\documentclass[usenatbib]{mnras}

\usepackage{float}
\usepackage{comment}

\usepackage[T1]{fontenc}

\DeclareRobustCommand{\VAN}[3]{#2}
\let\VANthebibliography\thebibliography
\def\thebibliography{\DeclareRobustCommand{\VAN}[3]{##3}\VANthebibliography}

\usepackage{graphicx}	
\usepackage{amsmath}	
\usepackage{amssymb}
\usepackage{bm}

\usepackage{newtxtext,newtxmath}
\usepackage[dvipsnames]{xcolor}
\usepackage{breqn}

\title[Foreground Map Errors in Global 21-cm Experiments]{A General Bayesian Framework to Account for Foreground Map Errors in Global 21-cm Experiments}

\author[M.~Pagano et al.]{
Michael Pagano,$^{1}$\thanks{E-mail: michael.pagano@mail.mcgill.ca}
Peter Sims$^{1}$\thanks{E-mail: psims@physics.mcgill.ca}
{Adrian Liu$^1$},
{Dominic Anstey$^{2,3}$},
\newauthor
{Will Handley$^{2,3}$},
{Eloy  de~Lera~Acedo$^{2,3}$}
\\
$^{1}$Department of Physics and McGill Space Institute, McGill University, Montreal, QC, Canada H3A 2T8\\
$^{2}$ Cavendish Astrophysics, University of Cambridge, Cambridge, UK\\
$^{3}$Kavli Institute for Cosmology, Madingley Road, Cambridge CB3 0HA, UK\\
}
\date{Submitted May 21st, 2020}

\begin{document}
\label{firstpage}
\pagerange{\pageref{firstpage}--\pageref{lastpage}}
\maketitle
\begin{abstract}
Measurement of the global 21-cm signal during Cosmic Dawn (CD) and the Epoch of Reionization (EoR) is made difficult by bright foreground emission which is 2-5 orders of magnitude larger than the expected signal. Fitting for a physics-motivated parametric forward model of the data within a Bayesian framework provides a robust means to separate the signal from the foregrounds, given sufficient information about the instrument and sky. It has previously been demonstrated that, within such a modelling framework, a foreground model of sufficient fidelity can be generated by dividing the sky into $N$ regions and scaling a base map assuming a distinct uniform spectral index in each region. Using the Radio Experiment for the Analysis of Cosmic Hydrogen (REACH) as our fiducial instrument, we show that, if unaccounted-for, amplitude errors in low-frequency radio maps used for our base map model will prevent recovery of the 21-cm signal within this framework, and that the level of bias in the recovered 21-cm signal is proportional to the amplitude and the correlation length of the base-map errors in the region. We introduce an updated foreground model that is capable of accounting for these measurement errors by fitting for a monopole offset and a set of spatially-dependent scale factors describing the ratio of the true and model sky temperatures, with the size of the set determined by Bayesian evidence-based model comparison. We show that our model is flexible enough to account for multiple foreground error scenarios allowing the 21-cm sky-averaged signal to be detected without bias from simulated observations with a smooth conical log spiral antenna.
\end{abstract}

\begin{keywords}
reionization, first stars -- large-scale structure of Universe -- methods: observational -- methods: statistical
\end{keywords}



\section{Introduction}
\label{sec:introduction}

    One of the remaining unmeasured periods in the history of our Universe is Cosmic Dawn (CD), the period in which the first stars are formed. These first generation stars give rise to the Epoch of Reionization (EoR), the stage in which the neutral hydrogen in the intergalactic medium (IGM) is ionized by the first galaxies. CD and EoR take place after formation of the Cosmic Microwave Background (CMB), but precede observations of the modern universe \citep{MoralesWyitheReview, Furlanetto2006Review, LoebFurlanetto2013, BarkanaLoeb2001}. The EoR and CD therefore represent the missing link between the early universe and the structure we see today. Despite its importance, a direct measurement of the EoR has been difficult due to a lack of direct probes. Thus far we have only indirect detection of its existence based on Ly$\alpha$ absorption and CMB optical depth measurements \citep{Planck}. 

    One of the most promising methods to make direct detection of CD and the EoR is to use the 21-cm hyperfine transition of Hydrogen in which a 21-cm wavelength photon is emitted when the electron flips its spin relative to the Hydrogen nucleus. Thus the 21-cm line directly probes the neutral hydrogen in the IGM during CD and the EoR. The emitted 21-cm wavelength photon is then redshifted according to cosmic expansion, which also allows for tomographic measurements of the neutral hydrogen along the line of sight. The 21-cm radiation falls in the radio portion of the electromagnetic spectrum and is measured in contrast to the CMB. The temperature difference between 21-cm photons and the CMB is referred to as the differential brightness temperature $\delta T_b$ \citep{Furlanetto2006Review, LiuShawReview2020}. 
    
    There have been two main approaches employed by experiments to measure $\delta T_b$ during the EoR. One approach uses interferometers which are sensitive to the spatial fluctuations of $\delta T_b$. This has been the approach of collaborations such as Hydrogen Epoch of Reionization Array (HERA, \citealt{HERA}), The Low-Frequency Array (LOFAR, \cite{LOFARintro}), The Murchison Widefield Array (MWA , \citealt{MWAintro}), Precision Array to Probe the Epoch of Reionization (PAPER, \citealt{PAPERintro}), and the Square Kilometre Array (SKA, \citealt{SKAintro}). An alternative approach to measuring $\delta T_b$ are 21-cm global signal measurements where the evolution of the spatially averaged $\delta_b$ as a function of frequency. This has been the approach of experiments such as Broadband Instrument for Global HydrOgen ReioNisation Signal (BIGHORNS, \citealt{BIGHORNSintro}), Experiment to Detect the Global EoR Signature (EDGES, \citealt{EDGESintro}), Large aperture Experiment to detect the Dark Ages (LEDA , \citealt{LEDAintro}),  Probing Radio Intensity at high-Z from Marion (PRIZM, \citealt{PRIZMintro}), The Giant Metrewave Radio Telescope (GMRT, \citealt{PacigaGMRT}), Radio Experiment for the Analysis of Cosmic Hydrogen (REACH, \citealt{REACH}), and Shaped Antennas to measure the background RAdio Spectrum (SARAS, \citealt{SARASFirstResults}). In this case, global signal experiments measure the mean of the differential brightness temperature, i.e, $T_b \equiv \overline{\delta T}_b$ given by:

\begin{eqnarray}
\label{eq:dTb}
     T_b( z) \!\! &\approx& \!\!\! 23 \left [ 1 - \overline{x}_{\rm HII}\right ] \left (  \frac{T_s(z) - T_{\rm CMB}(z)}{T_s(z)}  \right ) \left ( \frac{\Omega_b h^2}{0.023} \right)\\ 
     &\phantom{\times} &\times \left [  \left (  \frac{0.15}{\Omega_m h^2}  \right ) \left (  \frac{1 + z }{10}  \right )  \right ]^{1/2}  \textrm{mK}
\end{eqnarray}
where $\overline{x}_{\rm HII}$ is the mean ionized fraction of hydrogen,  $H(z)$ is the Hubble parameter (with $h$ as its dimensionless counterpart), $T_{\rm CMB}(z)$ is the CMB temperature, $\Omega_m$ is the normalized matter density, $\Omega_b$ is the normalized baryon density, and where $z$ is the redshift related to the frequency $\nu$ of the redshifted 21-cm photons by $1+z = \nu/\nu_0$ where $\nu_0$ is the rest frequency of a 21-cm photon. The mean differential brightness temperature (hereafter referred to as the global signal) depends on the spin temperature $T_s(z)$ of the neutral hydrogen gas, which measures the relative number of HI atoms that are in the excited versus ground hyperfine states. 
    
    The global 21-cm signal is an important probe of CD and the EoR. After the time of recombination, $T_s$ and the neutral hydrogen gas temperature $T_k$ are coupled due to collisional excitations. The remaining free electrons from recombination scatter off CMB photons keeping $T_{\rm CMB}$ coupled with $T_s$ leading to no 21-cm signal. After redshifts of $z \sim 150$ $T_s$ and $T_k$ decouple from $T_{\rm CMB}$. As the Universe expands the temperature of the neutral hydrogen gas $T_k$ cools adiabatically. Thus due to collsional coupling and cooling gas temperature, the spin temperature $T_s$ drops along with the gas temperature $T_k$ leading to the dark ages absorption trough. As the Universe expands collisions become more infrequent and eventually $T_k$ decouples with $T_s$ which once again falls into thermal equilibrium with $T_{\rm CMB}$. Once CD is underway, the first generation stars begin emitting Ly$\alpha$ photons which again couple $T_s$ to $T_k$ through the Wouthuysen-Field effect \citep{WFieldEffect52, Field58}. The spin temperature is again driven to the gas temperature $T_k$ which creates an absorption signature relative to $T_{\rm CMB}$. The redshift, duration and amplitude of the absorption trough due to the Ly$\alpha$ coupling directly depends on the details of reionization. Detection of this absorption trough was first reported by EDGES \citep{EDGESDetection}. However, the unexpectedly large amplitude and flattened profile was considerably larger than was predicted by astrophysical and cosmological models \citep{ReisFialkovBarkanaLya}. This has spurred theoretical interest in explaining the profile with excess radio background models, millicharged dark matter models and non-$\Lambda$CDM cosmology \citep{AEWRadioBG, FialkobBarkanaExcessRadio, BarkanaDMEDGES,HillDarkEnergyEDGES}. Other works have shown that the profile may be the result of an unaccounted for systematic \citep{SaurabhEDGESnonDetection,HillsEDGESDoubt,SaurabhEDGESdoubt,BradelyGroundPlaneEDGES,SimsPoberEDGES, MaxSmoothHarry}. As has been discussed by \cite{Tauscher2020, MartaSpinelliBeam, DominicInstrument, SimsPoberEDGES} improper modeling of the beam in the analysis of the data is one route through which such a systematic can be introduced. Finally as the first generation stars begin emitting x-rays, the gas temperature is heated, causing $T_s$ to create an emission spectrum relative to $T_{\rm CMB}$ \citep{PritchardLoeb2012, LiuParsons, JordanGlobalReview, JordanLamarre, Furlanetto, FialkovBerkanaGlobal, CohenFialkov, AdrianPritchardDesigner, BernardiGlobal2016, HarkerGlobalSignalMCMC}. Clearly, much of the physics of CD and EoR is encoded in precision measurements of the global signal.
    
    One of the difficulties in detecting the global 21-cm signal are the radio foregrounds (hereafter referred to as just foregrounds) which are 2-5 orders of magnitude brighter than the expected 21-cm signal within the expected redshift range of CD and the EoR. The effect of the foregrounds on global signal experiments has been well characterized in previous works \citep{BernardiFGPartI, BernardiFGPartII, FGPointSourceContAdrianMaxZal, ShaverFGDifficult}. The foregrounds are mostly due to synchrotron radiation which cause them to be smooth as a function of frequency which have allowed other studies to take advantage of their spectral smoothness to extract the 21-cm signal \citep{FGfitInFourierAdrianMaxJuddMatias}. Several techniques employ a model and subtraction strategy where the global signal is extracted by removing the foregrounds \citep{LeronGleserFGWeinerFilterSubtract}. In general there are a diverse set of foreground mitigation techniques that have been well developed. \citep{TauscherFGextract1_2018, TauscherFGextract2_2020, TauscherFGextract3_2020, TauscherFGextract42021, HarkerNonParametricSubtraction, JuddBowmanFGsubtract, PeterEDGES2023}.
    
    The Radio Experiment for the Analysis of Cosmic Hydrogen (REACH) has embraced a forward modelling approach to mitigate the foregrounds in order to detect the global 21-cm signal. A forward model approach requires modeling the instrument, the atmospheric environment, as well as the individual components of the radio sky, i.e, the galactic and extragalactic radio foregrounds and the cosmic 21-cm signal \citep{DominicInstrument, REACHSpectralModel, EmmaDominicIonosphere}. These radio components and all relevant errors are propagated though the analysis. Forward modelling the foregrounds requires modeling the spatial distribution of the foregrounds as well as their spectral distribution. This was previously explored in \cite{REACHSpectralModel}. To model the spatial distribution of the foregrounds, \cite{REACHSpectralModel} used radio sky models such as \cite{150MHz_AllSky, Haslam408, GSM2008, Zheng2017RadioSky}. These radio sky maps were measured
    without detailed error maps, rather only with quoted uncertainties on the offset and scale errors in the map (for example at $150$MHz it has been shown by \cite{SARASFirstResults, EDGESDetection} that the uniform scale errors are at the $\sim$5\%-11\% level). Thus, forward modelled foregrounds
    relying on these maps will contain spatial temperature perturbations (or spatial ``amplitude'' perturbations) in the
    forward model relative to the dataset that is observed by the instrument. In this paper we extend the foreground model in \cite{REACHSpectralModel} to account for the spatial errors in the foregrounds.

    In this paper we build on the work of \cite{REACHSpectralModel} and show that perturbations in the amplitude of these radio maps can cause systematics in the analysis of global 21-cm signal experiments which if unaccounted for will bias or make recovery of the true signal impossible. As of yet there is no accepted error model for these radio sky maps. In this work we bracket the range of systematic scenarios by studying a bracketing range of error scenarios and morphologies. We also study the effect that temperature offsets in radio sky maps have on our analysis. We then introduce a framework by which we can account for the amplitude and offset errors within the analysis. We show that our framework can recover the 21-cm signal without bias. We perform the analysis of this framework using the REACH experiment as our fiducial instrument; however, this framework is applicable to any global 21-cm experiment. 
    
   This paper is structured as follows. In Section \ref{sec:REACH_introduction} we introduce our fiducial global 21-cm experiment REACH and discuss our forward model of the foregrounds as well as our procedure to create simulated datasets. In Section \ref{sec:ErrorScenarios} we discuss the range of possible error scenarios in radio sky maps and the systematics that they cause in global 21-cm analysis. We also introduce our most realistic foreground map error scenario and use it as our fiducial foreground error realization in our dataset. In Section \ref{sec:AmpltiudeScaleFactorsBigSection} we introduce the amplitude scale factor framework that we use to account for foreground amplitude systematics. In Section \ref{sec:IsolatedAmplitudeErrors} we apply this framework to simulated datasets containing our fiducial foreground errors but assuming perfect knowledge of the spectral structure of the foregrounds. In Section \ref{sec:ResultsSpectrallyComplex}  we again apply this framework to our simulated datasets containing our fiducial foreground errors but now also assume imperfect a priori knowledge of their spectral structure and jointly fit for this, in the manner described in \cite{REACHSpectralModel}. We then conclude in Section \ref{sec:Conclusion}.

\section{Fiducial Instrument and Analysis Pipeline}
\label{sec:REACH_introduction}

The Radio Experiment for the Analysis of Cosmic Hydrogen (REACH) \citep{REACH} is a global 21-cm signal experiment located in the South African Karoo Desert. REACH will have observation windows corresponding to frequencies 50MHz to 170MHz corresponding to redshifts $z \simeq 28 - 7$. This is the expected redshift range for CD and EoR signals. REACH forward models the 21-cm signal, antenna beam and the galactic foregrounds as a function of $\nu$. The observation data taken over a set of LSTs is then fit to a forward modelled sky plus an additional signal model component. In this section we describe our simulated datasets, foreground model and Bayesian analysis of the simulated datasets.

\subsection{Bayesian Evidence}
\label{sssec:BayesTheoremPolychord}
Given a model $M$ of the sky which is parameterized by a set of cosmological parameters $\boldsymbol \theta$, we can infer the probability distribution of $\boldsymbol \theta$ given the measured dataset $\bf{d}$ for our particular model i.e. $p( \boldsymbol \theta | \mathbf{d}, M  )$. This is the posterior distribution in Bayes theorem:

\begin{equation}
    p( \boldsymbol \theta  | \mathbf{d} , M ) = \frac{p( \mathbf{d} | \boldsymbol \theta, M  ) p( \boldsymbol \theta |  M )}{p(\mathbf{d} | M ) }
\end{equation}
where $p( \mathbf{d} | \theta , M  )$ is the likelihood, $p(\boldsymbol \theta | M )$ is the prior on the parameter set $\theta$ and $p(\mathbf{d} | M)$ is the Bayesian Evidence. Throughout this paper we denote the Bayesian evidence as $Z$. The Bayesian Evidence is the normalization factor which ensures that the posterior of Bayes' theorem is a properly normalized probability distribution function. To numerically estimate the posterior probability distributions $p( \boldsymbol \theta | \mathbf{d},  M )$ of the parameters, one draws samples from the prior distribution $p( \boldsymbol \theta | M )$. Bayes' theorem can also be used for model selection, where one can determine the relative probabilities of models for the data via calculation of the Bayesian evidence,
\begin{equation}
    \label{eq:BruteForceEvidence}
    p(\mathbf{d} | M ) = \int d\boldsymbol \theta p( \mathbf{d} | \boldsymbol \theta,  M  ) p( \boldsymbol \theta | M )
\end{equation}

In order to compute the Bayesian evidence for each model we used the nested sampler \textsc{PolyChord} \citep{PolychordWill1, PolychordWill2, SkillingNestedSampler} which greatly decreases the computational cost required to compute the Bayesian Evidence. \textsc{PolyChord} operates by placing $n_{\rm live}$ points within the prior volume and then slice sampling the prior subject to the constraint that the new point has a higher likelihood than the lowest likelihood point. That lowest point is then discarded and the process repeats, generating new live points at each iteration. The sampling occurs from regions of low likelihood to high likelihood (see \cite{PolychordWill1, PolychordWill2} for more details). The Bayesian evidence is implicitly computed as the sampler gradually makes its way to the region of high likelihood. This sampling technique is also more effective at sampling multi-modal posteriors as compared to Markov chain Monte Carlo (MCMC) methods. The relevant \textsc{PolyChord} parameters for our analysis are n$_{\rm prior}$, the number of samples drawn from the prior, n$_{\rm fail}$ the number of consecutive failed sample draws at which the algorithm should terminate, and $\rm{n}_{\rm repeats}$ which toggles the number of repeats of the slice sampling procedure. Throughout our analysis we use the default values for these parameters, $n_{\rm live} = 25 n_{\rm dim}$ , $n_{\rm prior} = 25 n_{\rm dim}$ , $n_{\rm fail} = 25 n_{\rm dim}$ and $n_{\rm repeats} = 5 n_{\rm dim}$ where $n_{\rm dim}$ are the number of parameters in the inference. 

\begin{figure}
    \label{fig:spectral_map}
    \includegraphics[width=8cm]{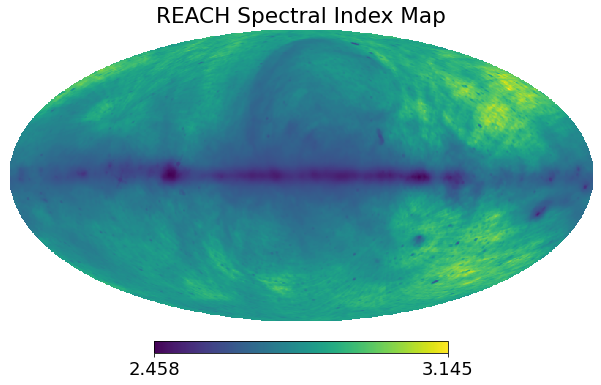}
    \caption{The spatial variation of the spectral index $\beta$  shown in galactic coordinates. Bright yellow regions correspond to larger spectral indices while darker regions correspond to small spectral indices. The morphology of $\beta$ roughly follows the galactic morphology. }
    \label{fig:BetaMap}
\end{figure}

\begin{figure*}
    \includegraphics[width=0.9\textwidth]{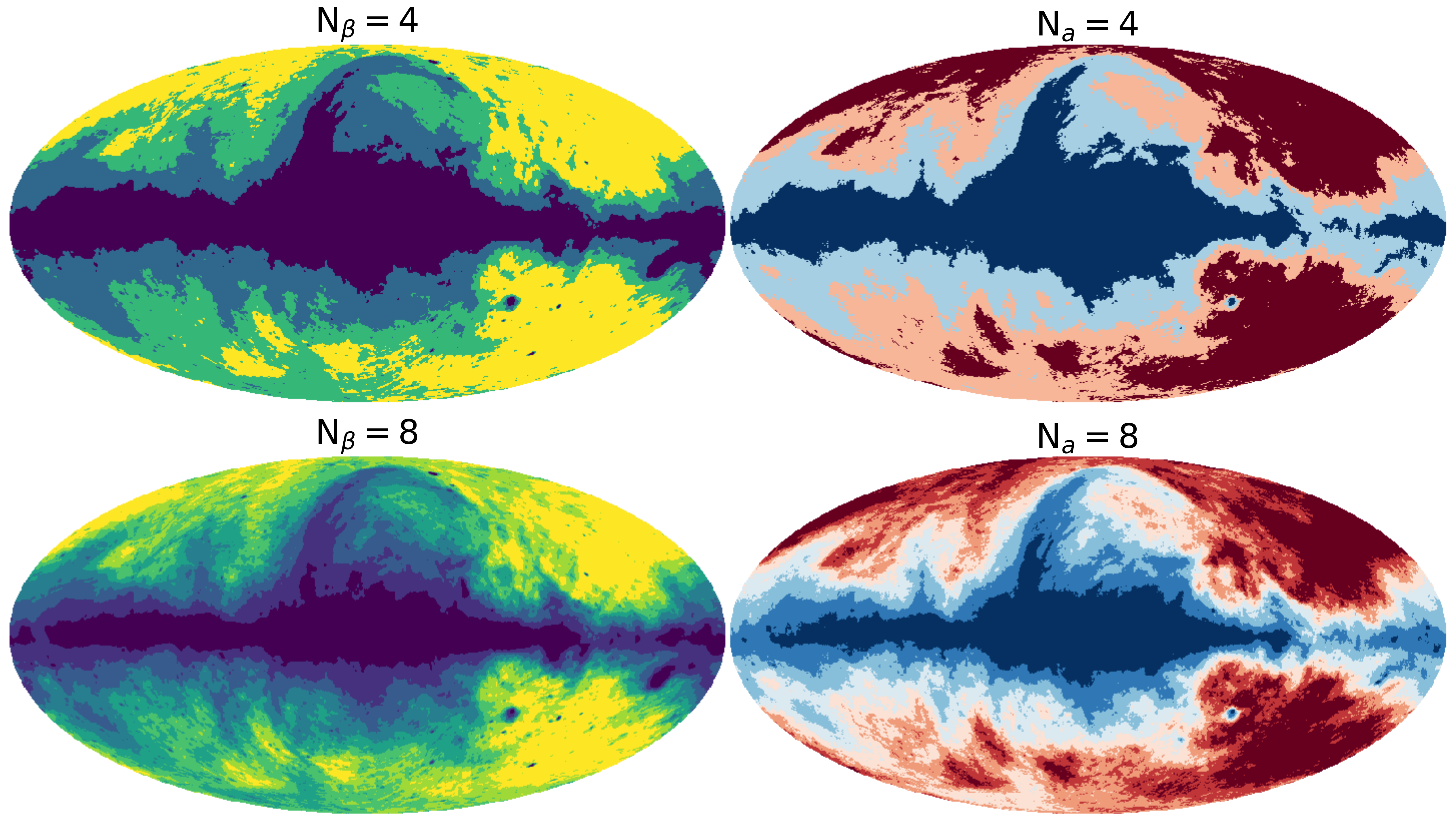}
    \caption{In each panel we show the division of the regions on the sky (displayed in Galactic coordinates). The left column corresponds to splitting the spectral map shown in Figure \ref{fig:BetaMap} into 4 regions (top) and 8 regions (bottom) (see \citealt{REACHSpectralModel}). The right column corresponds to amplitude scale factor regions (see Section \ref{sec:AmpltiudeScaleFactorsBigSection}). Here we split the Haslam map into 4 regions (top) and 8 regions (bottom).}   
    \label{fig:RegionsSplit}
\end{figure*}

\subsection{Data Simulations}
\label{sssec:DataSimulations}

In this section we describe our simulated dataset. The temperature of the sky within our observation band is modeled as the sum of the radio foregrounds, CMB and the cosmological signal $T_{\rm sky}(\theta,\phi,\nu) = T_{\rm 21}(\nu) + T_{\rm FG}(\theta,\phi,\nu) + T_{\rm CMB}$ . The temperature of the radio foregrounds are dominated by synchrotron and free-free radiation emission which are individually well described by spatially-dependent spectral power laws. Following \cite{REACHSpectralModel}, we form a model for these foregrounds as,

\begin{equation}
\label{eq:TFG}
T_{\rm FG}(\theta,\phi, \nu) = (T_{\rm base}(\theta, \phi, \nu_0)  - T_{\rm CMB})(\nu/\nu_0)^{-\beta(\theta, \phi)}
\end{equation}
where $\beta(\theta, \phi)$ is the spatially varying spectral index and $\nu_0$ is the reference frequency of the foreground model which we set to $408$MHz. A map of spectral index variation across the sky was previously derived in \cite{REACHSpectralModel} by calculating the spectral index required to map each pixel of the Global Sky Model (GSM, \citealt{GSM2008}) at $230$MHz to the corresponding pixel of the GSM map at 408MHz, i.e.
\begin{equation}
    \beta(\theta, \phi) = \frac{\textrm{ln}\left(\frac{ T_{230}(\theta,\phi) - T_{\gamma}}{T_{408}(\theta,\phi) - T_{\rm CMB}} \right) }{\textrm{ln}(\textrm{230}/\textrm{408})}
\end{equation}
The resulting spectral index map is shown in Figure \ref{fig:BetaMap}.
$T_{\rm base}$ is the temperature of the sky measured at a reference frequency $\nu_0$. This map, which we refer to as the  foreground ``basemap'' is then extrapolated in frequency according to $\beta(\theta,\phi)$. Since REACH is a drift scan experiment, the resulting temperature of the sky is then rotated for time and observation location of the instrument. The resulting dataset as seen through the beam is then
\begin{equation}
    \label{eq:DatasetAfterBeam}
    T_{\rm sky} = \frac{1}{4\pi}\int_0^{4\pi}D(\theta, \phi,\nu)\int_{\rm LST_0}^{\rm LST}T_{\rm FG}(\theta,\phi,\nu,t)dtd\Omega
\end{equation}
where $D(\theta, \phi,\nu)$ is the antenna's directivity at frequency $\nu$ and $\rm{LST}_0$, $\rm{LST}$ refer to the beginning and end of the observation times. Note that the REACH pipeline can also accommodate models where Equation \ref{eq:DatasetAfterBeam} is not time averaged (see \citealt{DominicLSTDependentPipeline}). We model the 21-cm signal as a simple Gaussian of amplitude $A_{\rm 21}$ , width $\sigma_{\rm 21}$ centered at frequency $\nu_{\rm 21}$: 
\begin{equation}
    \label{eq:21-cmGaussian}
T_{\rm 21} = A_{\rm 21} e^{ -\frac{(\nu - \nu_{\rm 21})^2}{2\sigma_{\rm 21}^2}  }
\end{equation}
This model of the 21-cm signal is deliberately
   phenomenological, providing enough flexibility to characterize the range of realistic scenarios in our signal recovery analysis. The cosmological signal $T_{\rm 21}$ and the CMB $T_{\gamma}$ are taken to be spatially uniform. The temperature of the sky is then computed as 
\begin{equation}
    \label{eq:DataSkyTemperatureFinal}
    T_{\rm data}(\nu) = T_{\rm sky}(\nu) + T_{\rm 21}(\nu) + T_{\rm CMB} +\sigma_{\rm noise}
\end{equation}
where $\sigma_{\rm noise}$ is random uncorrelated Gaussian noise with standard deviation $0.025$K. Note that we chose this noise model for better comparison with previous work such as \cite{REACHSpectralModel} which uses the same fiducial noise level. However the REACH pipeline is flexible enough to accommodate radiometric noise
\citep{Scheutwinkel2022}, and to model select across noise models \citep{Scheutwinkel2022B}.

\subsection{Spectral Sky Model}
\label{sssec:spectral_model}

In the previous section we introduced our fiducial map for the spatially varying spectral index which was used to create the dataset. Since the true spatial variation of the spectral index is not known, the spectral indices must be fit for within our foreground model. This spectral model was the main focus of \cite{REACHSpectralModel} which we briefly describe here.  Since fitting for the spectral index at each pixel on the sky is computationally infeasible, in order to model the spatial variation of the spectral index in a computational feasible manner, the spectral map is split into $N_\beta$ regions of uniform $\beta$. The regions are selected such that the spectral indices are similar within the region. This framework becomes a closer approximation to the true spectral sky as the number of regions $N_\beta$ is increased. Our foreground model can then be written as

\begin{equation}
\label{eq:REACHForegroundModelSpectralMasks}
T_{\rm FG}(\theta,\phi, \nu)  = \left [ \sum_{i}^{N_\beta} M_{\beta,i}(\theta,\phi) (T_{\rm base}(\theta, \phi, \nu_0)  - T_{\rm CMB})(\nu/\nu_0)^{-\beta_i} \right]
\end{equation}
where $i$ are the indicies corresponding to $i$th spectral region.
The values $M_{\beta,i}(\theta, \phi)$ are masks that take on the value of 0 or 1 thereby determining whether a pixel is part of the $i$th spectral region which contains a spectral index $\beta_i$. On the left column of Figure \ref{fig:RegionsSplit} we show example regions where the spectral map from Figure \ref{fig:BetaMap} has been split into $4$ regions (top) and $8$ regions (bottom). The foregrounds are then rotated onto the coordinates of the beam at the specified LSTs (corresponding to the observational times). The sky regions as measured by the instrument are then computed as
\begin{equation}
    \label{eq:SkyModelAfterBeam}
    T^{i}_{\rm{sky}}(\nu) = \frac{1}{4\pi}\int_0^{4\pi}D(\theta, \phi,\nu)\int_{\rm{LST}_0}^{\rm LST}T^{i}_{{\rm FG}}(\theta,\phi,\nu,N_\beta,t)dtd\Omega .
\end{equation}
where $T^{i}_{\rm{sky}}$ is the contribution from the $i$th region of the sky to the measured spectrum. The mean temperature of the sky is then computed as
\begin{equation}
    \label{eq:TempModelOFSkyFinal}
    T_{\rm model}(\nu) = \sum_i^{N_\beta} T^{i}_{\rm{sky}}(\nu) + T_{\rm 21}(A_{\rm 21}, \nu, \sigma_{\rm 21} )+ T_{\rm CMB}
\end{equation}
where $ T_{\rm 21}(A_{\rm 21}, \nu, \sigma_{\rm 21} )$ refers to the model 21-cm described by Equation \ref{eq:21-cmGaussian} with amplitude, central frequency and width  $(A_{\rm 21}, \nu_{\rm 21}, \sigma_{\rm 21} )$. We compare the sky model $T_{\rm sky}$ to the simulated dataset $T_{\rm data}$ through the Gaussian likelihood
\begin{equation}
\label{eq:LikelihoodTimeAveraged}
\textrm{ln}L = -\frac{1}{2} \sum_j \textrm{ln}\left ( 2\pi \sigma_n^2 \right ) - \frac{1}{2} \left (  \frac{T_{\textrm{ data}}(\nu_j) - T_{\textrm{model} }(\nu_j)      }{\sigma_n^2}		\right )^2   
\end{equation}
where the index $j$ refers to the frequency bin. We assume that the noise is uncorrelated between frequency bins. Throughout this paper we place a uniform prior on the spectral indices $\beta$ between [2.45, 3.15] which encompasses the entire range of spectral indices in the spectral index map in Figure \ref{fig:BetaMap}. Further we place uniform priors on the 21-cm signal parameters $(A_{\rm 21} , \nu_{\rm 21} , \sigma_{\rm 21}): [0, 0.25], [50$MHz$, 150$MHz$], [10$MHz$, 20$MHz$]$. In principle, we could jointly estimate the sky model and noise in the data as demonstrated in \cite{REACHSpectralModel}; however, for computational simplicity, here we fix the noise estimate at  $\sigma = 0.025$K. Unless otherwise stated, we use a $1$ hour observation time starting at $18:00:00$ on $2021-01-01$. We use a conical-log spiral antenna as our fiducial instrument (located in the South African Karoo desert) throughout our analysis. In this work we also assume that the beam model is precisely known, i.e $D$ in Equation \ref{eq:DatasetAfterBeam} is equal to $D$ in Equation \ref{eq:SkyModelAfterBeam}. In practice the uncertainties in the beam model are also liable to introduce additional instrumental chromaticity (e.g. \citep{PeterPober2020, MartaSpinelliBeam}) which can bias recovery of the 21-cm signal. Accounting for beam uncertainty in the context of the analysis pipeline presented here will be explored in future work (Anstey et al in prep).

\section{Foreground Errors}
\label{sec:ErrorScenarios}

Detailed foreground sky maps such as those derived from the Haslam sky survey at 408 MHz (\citealt{Haslam408}), the \cite{150MHz_AllSky} sky survey at 150MHz, and the \cite{Guzman45} sky survey at 45MHz are essential tools in analyses aimed at detecting redshifted 21-cm emission. Historically, sky surveys have not been accompanied by detailed error models characterizing the full covariance information in the map, which complicates detailed statistical forward modelling of the signals. Subsequent analyses comparing these maps to absolutely calibrated spectrometer data have sought to characterize their offset and scale factor uncertainties (e.g. \cite{PatraSARASoffset} and \cite{RaulEDGESoffset}). Additionally, recent surveys such as the LWA1 Low Frequency Sky Survey between 35 and 80 MHz \citep{Dowell2017} and EDA2 map of the southern sky at 159 MHz \citep{Kriele2022} included more detailed uncertainty characterization, but do not have full-sky coverage and are lower in resolution than the Haslam sky survey at 408MHz.

We will show in Section \ref{sec:ApplicationToSpectrallUniform} that the correlation length (which here we equate with map resolution) of the uncertainties in the base map used in our forward model strongly affects reconstruction accuracy for a given complexity of foreground model. In the context of modelling the 21-cm global signal, less regions are required to model the foregrounds at a level sufficient for unbiased 21-cm signal recovery when the errors have shorter correlation lengths, and vice versa. Thus, increasing the correlation length of the uncertainties in the base map have the effect of requiring more degrees of freedom to model the foregrounds accurately. As such, here, we make use of the $\sim 1^\circ$ resolution Haslam sky survey at 408MHz as our fiducial base model $T_{\rm base}$. We then consider multiple error scenarios $T_{\rm error}$, bracketing the possible systematic errors caused by amplitude perturbations in the foreground maps. This allows us to study the systematics created in the analysis in each of these cases and characterize the dependence of the model complexity on both foreground error level and correlation length (see Section \ref{sec:ApplicationToSpectrallUniform} for details). 
In Section \ref{sssec:ErrorAssessmentMethodology} we describe our method to assess whether the foreground error realisations produce significant systematics in our analysis. In Section \ref{sssec:other_error_scenarios} we build intuition on different foreground error scenarios, while in Section \ref{sssec:RealisticErrors} we introduce our method for generating more realistic foreground amplitude errors. 

\subsection{Methodology For Assessing Foreground Errors in the Analysis}
\label{sssec:ErrorAssessmentMethodology}
To assess whether a particular foreground error scenario produces significant systematics in our analysis, we create a simulated dataset using the procedure described in Section \ref{sssec:DataSimulations} but with our noise realization added to the basemap in Equation \ref{eq:TFG}, i.e. $T_{\rm base}^{\rm true}(\theta, \phi) = T_{\rm base}(\theta, \phi) +T_{\rm error}(\theta,\phi)$ where $T_{\rm error}(\theta, \phi)$ is the temperature map of the noise realization and $T_{\rm base}^{\rm true}$ is the true temperature of the sky.
To isolate the effect that the temperature perturbation errors have on the data analysis pipeline, we do not add 21-cm signal in the dataset. Note that for brevity we will refer to the spatial temperature perturbations in the basemap as amplitude errors. This condition essentially makes our assessment of the foreground errors a null test, i.e. there is no 21-cm signal inside the dataset, and so any recovered 21-cm signal is thus due to a systematic caused by the foreground amplitude errors. Thus the null test is passed if the amplitude of the recovered 21-cm signal is consistent with zero at $1\sigma$. Note that  one can also use the Bayesian evidence to test whether a signal is preferred. This is done by comparing the Bayesian
evidence for a model with a signal component versus one without a signal component. Our dataset is written as 
\begin{equation}
T_{\rm FG}(\theta,\phi, \nu)  =  ( T^{\rm true}_{\rm base}(\theta,\phi) - T_{\rm CMB})(\nu/\nu_0)^{-\beta} .
\end{equation}
The remainder of the simulated dataset procedure remains the same as discussed in Section \ref{sssec:DataSimulations}. The simulated dataset includes the estimated noise realization while the model does not include noise.
In our foreground model we fix the number of spectral regions to $N_{\beta} = 10$ which (for this antenna and observation settings) has been shown by \cite{REACHSpectralModel} to lead to unbiased recovery of the 21-cm signal. Thus $N_\beta$ provides sufficient flexibility to account for the spatial variation of the spectral index in the sky. By fixing $N_\beta = 10$, we avoid adding additional spectral components in the foreground model to fit for amplitude perturbations in the foreground basemap \footnote{Note that in Section \ref{sec:ResultsSpectrallyComplex} we show that spectral components cannot compensate for systematics that are caused by amplitude perturbations in the basemap.}. This assumption allows us to better isolate the effect of the amplitude errors in the analysis. Note that we only fix $N_\beta = 10$ in this section.  In the following section we consider a range of scenarios for $T_{\rm error}(\theta, \phi)$ and using this methodology study their affect on the analysis.


\subsection{Error Scenarios}
\label{sssec:other_error_scenarios}
Since we do not precisely know the nature or morphological structure of the errors present in the foreground map, we consider a range of error scenarios and amplitudes with the aim of bracketing the possibilities for the systematics.  Roughly, the errors we have considered fall into two categories :
\begin{enumerate}
    \item Noise Scenario 1: Homoscedastic errors
    \item Noise Scenario 2: Heteroscedastic errors
\end{enumerate}
Scenario 1 corresponds to error realisations where the standard deviation of the errors is uniform over the sky. In Scenario 2 we consider the case where the standard deviation of the errors are proportional to the temperature in each pixel of the basemap. In each scenario, we assess the systematics produced by the presence of these errors in the analysis through the methodology and observation settings discussed in Section \ref{sssec:ErrorAssessmentMethodology}. To produce the errors in Scenario 1, we add random Gaussian noise with mean $\mu = 0$ and of standard deviation $\sigma_{\rm noise} = \Delta \cdot \overline{\delta T}_{\rm sky}$ where $\overline{\delta T}_{\rm sky}$ is the mean temperature of the map and the $\Delta$ is a dimensionless parameter which we vary $\left[0.01, 0.05, 0.1\right]$ corresponding to random Gaussian error fluctuations from $1\%$ to $10\%$ of the mean temperature in the map. To imprint correlation structures into our error realisation  we smooth over the map by convolving the error realisation with a Gaussian symmetric beam. By modifying the FWHM of the Gaussian beam we can produce error realisations which are correlated on a physical length scale measured in radians on the sky $\theta_{\rm FWHM}$. We consider noise correlation scales corresponding to the full width half max (FWHM) of $\theta_{\rm FWHM} = \left[0, 1^\circ, 5^\circ, 10^\circ, \infty \right]$. The extreme correlation scenarios $\theta_{\rm FWHM} \rightarrow \infty$ and $\theta_{\rm FWHM} \rightarrow 0$  corresponds to a uniform offset in the map and uncorrelated random Gaussian noise respectively.
\begin{figure}
  \includegraphics[width=0.5\textwidth]{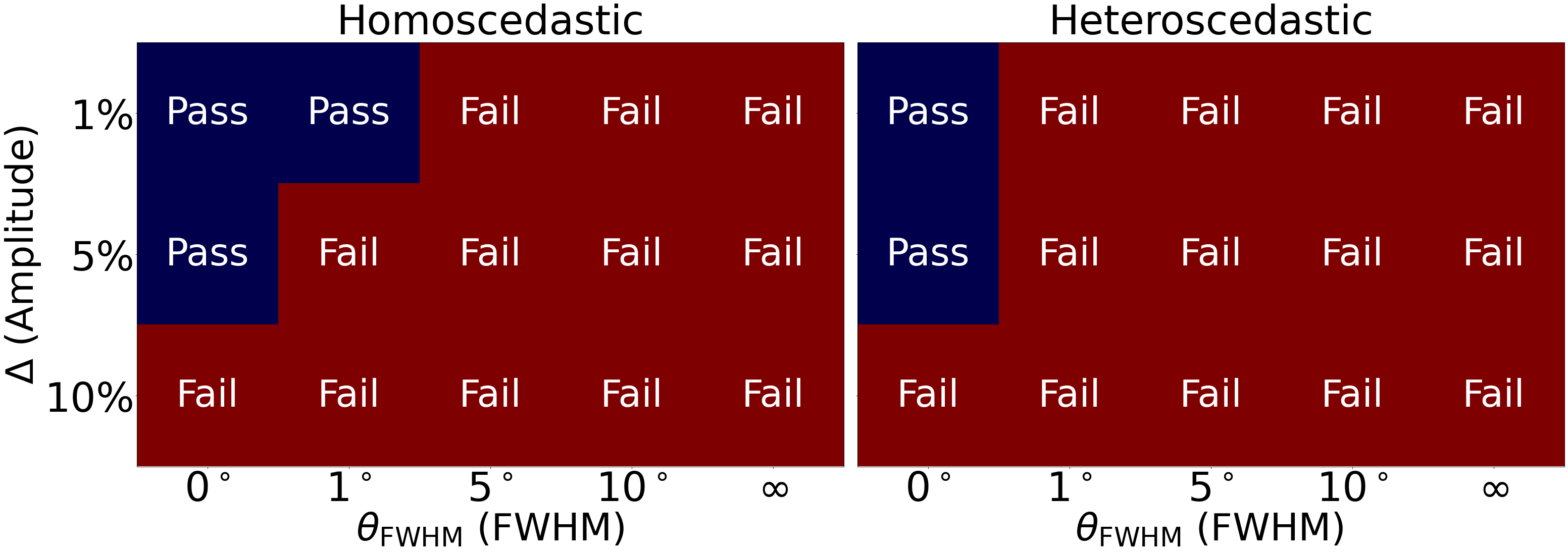}
  \caption{Summary of the null test results described in Section \ref{sssec:ErrorAssessmentMethodology}  for the error scenarios described in Section \ref{sssec:other_error_scenarios}. The horizontal axis corresponds to the FWHM of the error correlation length while the vertical axis corresponds to the amplitude of the fluctuations. A red panel indicates that the error scenario does lead to a systematic in the analysis large enough such that the null test fails. A blue panel indicates that the error scenario does not lead to a systematic large enough such that the null test fails. The left panel are error scenarios where the standard deviation of the errors are uniform across the sky while the right panel are error scenarios that have standard deviation which is spatially dependent. }
  \label{fig:PassFail}
\end{figure}

\begin{figure}
  \includegraphics[width=0.5\textwidth]{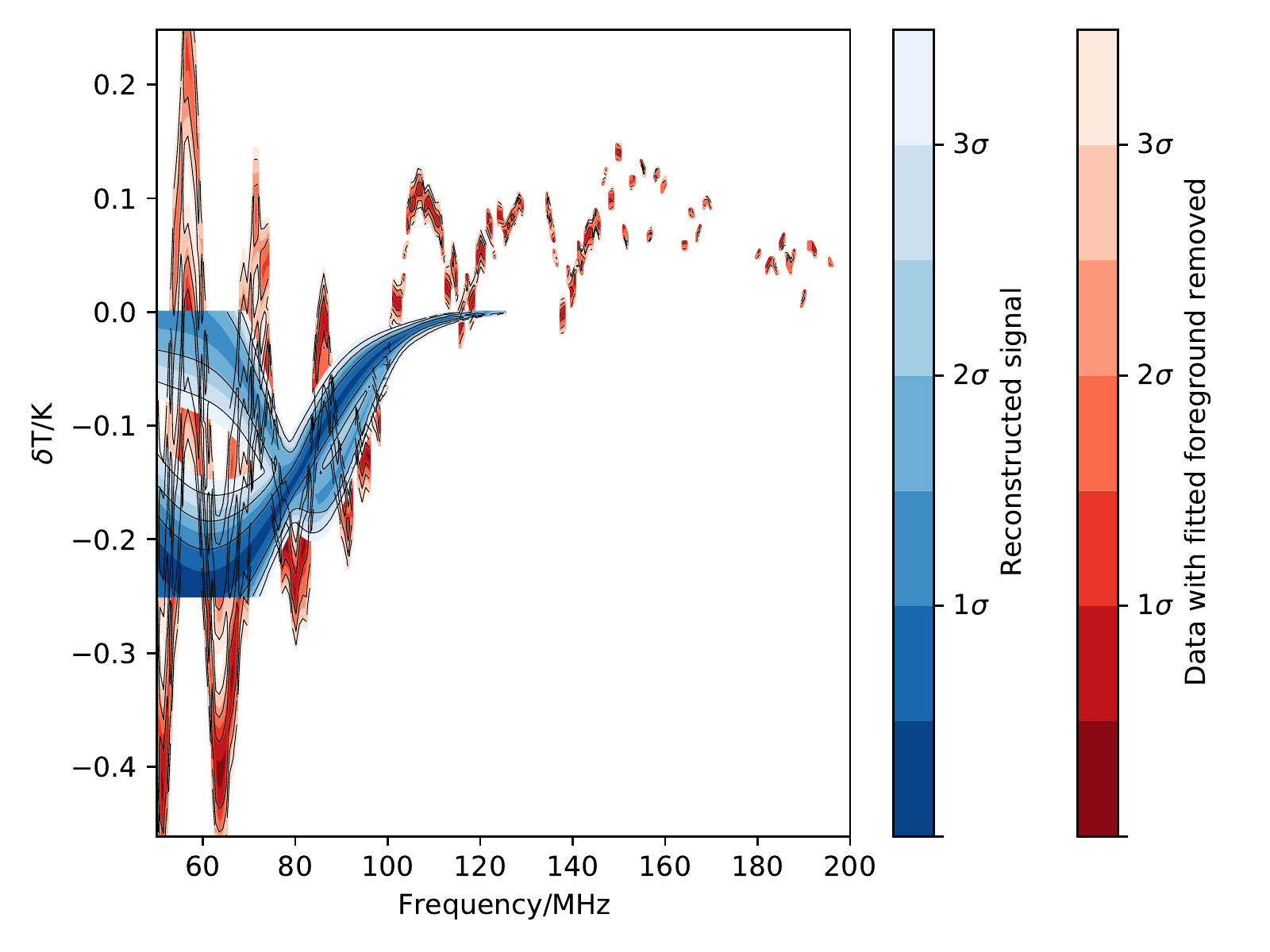}
  \caption{Signal recovery plot for the heteroscedastic error realisation with $\Delta = 5\%$ and $\theta_{\rm FWHM} = 1^\circ$. The blue curve represents the reconstructed 21-cm signal using Equation \ref{eq:21-cmGaussian} while the red curve is the dataset (Equation \ref{eq:DataSkyTemperatureFinal}) with the fitted foreground removed, i.e. Equation \ref{eq:REACHForegroundModelSpectralMasks} with the best fitting foreground parameters. The lightly shaded red and blue regions represent the $3\sigma$ regions for their respective contours. } The systematic produced by the errors in the basemap prevent the model from passing the null test described in Section \ref{sssec:ErrorAssessmentMethodology}
  \label{fig:SystematicErrorGrid}
\end{figure}

In the left panel of Figure \ref{fig:PassFail} we summarize the results for error Scenario 1. In each noise scenario we indicate whether the error scenario has passed or failed the null test according to our criteria above.  We find that uncorrelated Gaussian errors with fluctuations up to $5\%$ the mean temperature of the map do not produce significant enough systematics for our analysis to fail the null test described in Section \ref{sssec:ErrorAssessmentMethodology}. As we progressively increase the smoothing scale of the errors from $\theta_{\rm FWHM} = 0$ to $\theta_{\rm FWHM} = 10^\circ$, we find that the amplitude of systematics introduced into the data increases. In the limit that the errors are smoothed with $\theta_{\rm FWHM} \rightarrow \infty$ (i.e. a uniform offset) then even errors at $1\%$ produce a sufficiently large systematic for the null test to fail for the $25$mK noise level in the data considered here.

In Scenario 2 we consider the heteroscedastic version of Scenario 1 in which the random Gaussian noise have spatially dependent fluctuations, i.e. $\sigma_{\rm noise}$ acquires a spatial dependence. We again produce a range of error scenarios with temperature fluctuations $\sigma_{\rm noise} = \Delta \cdot \delta T_{\rm sky}(\theta, \phi)$ where $T_{\rm sky}(\theta, \phi)$ is the temperature of the Haslam at location $\theta, \phi$ on the sky and $\Delta$ is a unitless parameter which obtains the same values as Scenario 1. For each value of $\sigma_{\rm noise}$ we progressively smooth the map by convolving the errors with a Gaussian beam with FWHM $\theta_{\rm FWHM} = \left[0, 1^\circ, 5^\circ, 10^\circ, \infty \right]$. On the right panel of Figure \ref{fig:PassFail} we summarize the results of these tests. Qualitatively, we find that increasing the amplitude of the error fluctuations $\Delta$ or the correlation length $\theta_{\rm FWHM}$ of the errors produce larger systematics in the analysis. Quantitatively, we find that the systematic created by the heteroskedastic noise realization with amplitude $\Delta > 5\%$ and correlation length $\theta_{\rm FWHM} \ge 1^\circ$ is large enough to prevent recovery of the 21-cm signal. In Figure \ref{fig:SystematicErrorGrid} we show an example of a systematic created by the scenario $\Delta = 5\% $, $\theta_{\rm FWHM} = 1^\circ$. The blue curve represents the reconstructed 21-cm signal using the 21-cm model parameters described in Equation \ref{eq:21-cmGaussian} while the red curves represent the dataset (see Equation \ref{eq:DataSkyTemperatureFinal}) with the fitted foreground removed (Equation \ref{eq:REACHForegroundModelSpectralMasks}). The red curves can thus be interpreted as containing any 21-cm signal and residuals remaining after the fit. In an ideal scenario for the null test that we are conducting in this section, the blue curves are consistent with the true 21-cm signal (which is zero in this case since it is a null test) and red contours are reduced to the noise level of the dataset. Also note that the lighter blue and red parts of the plot represent the $3\sigma$ level of their respective contours. For fixed $\Delta$, increasing $\theta_{\rm FWHM}$ increases the likelihood that the null test fails. When $\theta_{\rm FWHM} \ge 1^\circ$, all error scenarios lead to systematics such that the null test outlined in Section \ref{sssec:ErrorAssessmentMethodology} are not passed. The largest systematics are caused by scenarios with $\theta_{\rm FWHM} = \infty$ and $\Delta = 10\%$. 
We further test this error scenario with longer observation settings (6hr integration time) as well as observation times where the galactic plane is predominantly below the horizon. The systematics caused by this type of error scenario are present in all observation configurations but are lower for longer integration times and when the galaxy is below the horizon.


We consider the errors generated in scenario 2 to be most realistic, i.e. we make the assumption that the fluctuations of the errors depend on the local temperature. We also assume that the errors have structure with some unknown correlation length $\theta_{\rm FWHM}$. In the following section we build on the intuition developed in this section and introduce our fiducial error scenarios used in our analysis.

\subsection{Fiducial Error Models}
\label{sssec:RealisticErrors}

\begin{figure}
  \includegraphics[width=0.5\textwidth]{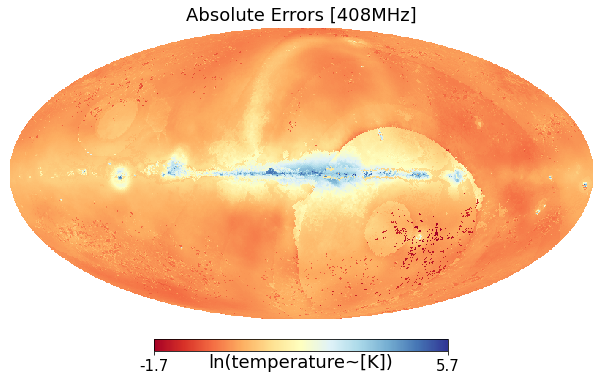}
  \caption{Absolute errors in the temperature of the sky at $408$MHz shown in Galactic coordinates. Note that the errors roughly trace the galactic morphology. Also apparent is that many low frequency survey datasets are missing the southern celestial pole which influences the eGSM model in this region.  }
  \label{fig:AveryAbsErrors}
\end{figure}

\begin{figure*}
  \includegraphics[width=0.95\textwidth]{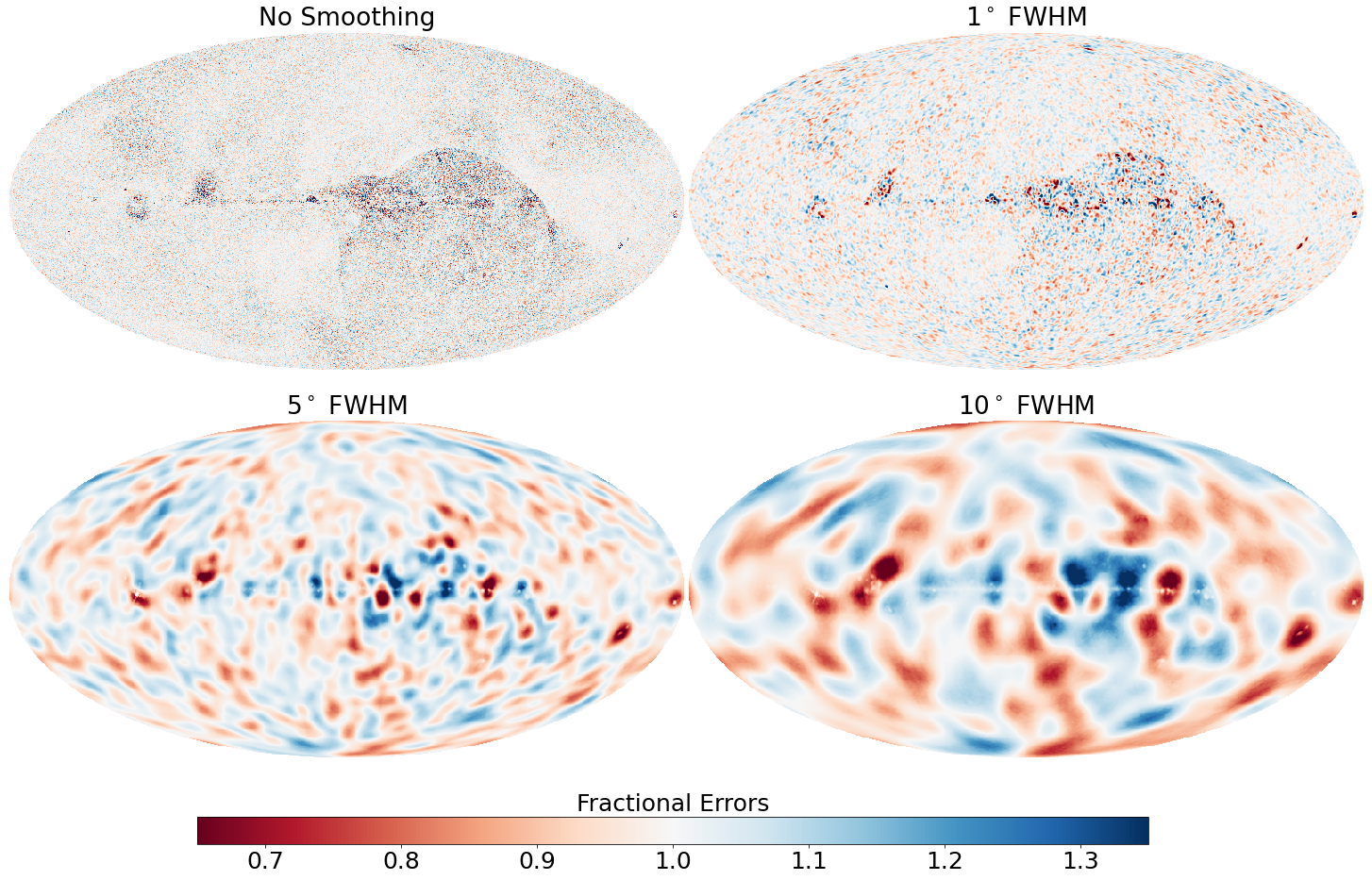}
  \caption{In this figure we show our fiducial error scenarios at $408$MHz. Each figure is displayed in Galactic coordinates. In the top left we show the a heteroskedatic error realization generated by drawing values from a Gaussian distribution of zero mean and standard deviation equal to the absolute errors at $408$MHz. Moving clockwise from the top left panel, the resulting error realisation is smoothed by a Gaussian beam of FWHM equal to $1^\circ$ (top right), $5^\circ$ (bottom left), and $10^\circ$ (bottom right). }
  \label{fig:FiducialErrorMap}
\end{figure*}

In the absence of an error structure for the radio foreground basemap, we need to construct our estimate for a realistic error scenario. Our construction of the errormap is motivated by the expectation that the errors likely have spatial structure, i.e. they are not entirely uncorrelated, and that the magnitude of the errors is proportional to the absolute temperature of the basemap. Thus our expectation is that the error fluctuations are largest in the galactic plane, where the temperature is the largest. As discussed in \cite{Haslam408} the absolute calibration of the Haslam 408MHz map is of the thought to be accurate to better than 10\% leading to an overall zero level of $\pm 3$K.  As a conservative estimate of the monopole error, here we assume that our map has an offset of $3\,\textrm{K}$. We do not expect our error realization to exactly match reality, however we use it as our fiducial error realisation to study the effect that such noise scenarios have on our data analysis pipeline.

We now discuss our procedure to generate our error realization. This is a two step process, the first step is to construct the spatial error fluctuations while the second step is to adjust the mean of the map to $3\,\textrm{K}$. To construct the spatial error fluctuations via a Gaussian generated heteroskedastic errormap with amplitude proportional to the temperature at each location in the sky, we use the extended Global Sky Model (eGSM). Specifically we use the eGSM absolute errors of the 408MHz Haslam map derived from Kim et al. (in prep) which we show in Figure \ref{fig:AveryAbsErrors}. The eGSM is a data-driven model similar in spirit to previous global sky models published by \cite{GSM2008} and \cite{Zheng2017RadioSky}, in that it takes empirical data from multiple surveys and interpolates over gaps in frequency and sky coverage using dimensional reduction methods such as principal component analyses. Importantly, the eGSM includes the Monte Carlo propagation of measurement errors in the input surveys through the interpolation process. The errors roughly trace the galactic morphology. Note that many low frequency survey data are missing the southern pole which influences the eGSM model in this region. Admittedly, the final error bars are only rough estimates given that many of the input surveys do not come with published uncertainties, thus making it necessary to simply \emph{assume} $\sim 10\% $ errors in some input maps. Nonetheless, Figure \ref{fig:AveryAbsErrors} likely captures some of the spatial patterns in the errors of typical sky models and is therefore sufficient for substantiating at least the qualitative lessons of this paper.

To go from the map of error bars in Figure \ref{fig:AveryAbsErrors} to an error \emph{realization}, we first draw a noise realization from the absolute error map at 408MHz. To do this, at each pixel in the basemap we draw a random value from a Gaussian distribution with mean $\mu = 0$ and standard deviation $\sigma_{\rm pixel}$ proportional to the value of the absolute error in that pixel. This produces a heteroscedastic noise realisation of zero mean and with error fluctuations proportional to the absolute errors at 408MHz. The fractional error fluctuations in our noise realization are largest in the galactic plane and smaller away from the galactic center. In the upper left of Figure \ref{fig:FiducialErrorMap} we show the resulting errormap. Note the areas of high foreground map uncertainty  in the southern pole (lower right portion of the basemap). Many low frequency survey data are missing the southern pole which influences the eGSM model in this region. 
In practice, with particularly poor empirical constraints in those parts of the sky, it would be sensible for experiments to avoid those regions. Note that this might cause further limitations for global signal experiments which are unable to avoid the southern pole. We intend to address this in future work. We select our observation time such that these regions are below the horizon at the time of observation. We chose to set our observation starting at 6pm on 2021-01-01. Unless otherwise stated, we use this observation time for the remainder of our analysis. 

To implement correlations into our noise realization we smooth the map by convolving our noise realization with a Gaussian beam of FWHM of $\theta_{\rm FWHM}$. Since we don't know the true correlation structure of the errors, we consider multiple smoothing scales chosen to encompass a range of physical scenarios. We use $\theta_{\rm FWHM} = \left [ 1 ^{\circ}, 5 ^{\circ}, 10 ^{\circ}\right ]$. These correlation scales are chosen to roughly correspond to the resolutions in the 150MHz and 408MHz empirical foregrounds maps from \cite{150MHz_AllSky, Haslam408} which are approximately $1^\circ$ and $5^\circ$ FWHM respectively. We make the assumption that the correlation lengths of the error features roughly match the resolutions of these maps. Note that smoothing the map obscures the features of the original eGSM map. For small smoothing scales the morphology of the underlying eGSM map is still obvious, however with increasingly large smoothing scales, the features of the underlying eGSM map are entirely obscured. In an extreme case when $\sigma_{\rm FWHM} \rightarrow \infty$, the errors are totally independent of the eGSM map (assuming they all have the same mean). Convolving our error realisation with a Gaussian beam reduces the total spatial power in the map. We wish to preserve the total power in the map. To quantify how the total power in the map is affected we first express the original (pre-smoothed) error map in spherical harmonics
\begin{equation}
    \label{eq:spherical_harmonics}
    Y_{lm} = \sqrt{\frac{2l + 1}{4\pi} \frac{(l-m)!}{ (l + m)! } } P_{l}^m (\cos(\theta)) e^{i m \phi}
\end{equation}
where the indicies $l = 0, ..., \infty$ are referred to the multipole which represent the angular scale on the sky, and $-l \le m \le +l$. The $P_l^m$ are the associated Legendre Polynomials. We can expand the map using 
\begin{equation}
    \label{eq:spherical_harmonics_expansion}
    \Theta( \hat{n}) = \sum_{l = 0}^{l = \infty} \sum_{m = -l}^{m = + l} a_{lm} Y_{lm}(\hat{n}) 
\end{equation}
where 
\begin{equation}
    \label{eq:alm}
    a_{lm} = \int \int \Theta( \hat{n}) Y^*_{lm}(\hat{n})d\Omega .
\end{equation}

We can define the power spectrum $C_\ell$ of the temperature fluctuations in the map which is the variance of the harmonic coefficients: 
\begin{equation}
    \label{eq:angular_pspec}
    \left < a_{lm}a^*_{l'm'} \right > = \delta_{l l'} \delta_{m m'}C_l
\end{equation}
where the angular brackets represent an ensemble average taken over many independent realizations in the map. If we perform the average over different $m$ modes we can write the angular power spectrum $C_l$ as 
\begin{equation}
    \label{eq:Cl_coefficients}
    C_l = \frac{1}{2l + 1} \sum^{m = +l}_{m = -l}\left <|a_{lm}|^2 \right > .
\end{equation}
The total power $\sum_{l}^{l_{\rm max}} C_l $ of the smoothed error map will have decreased power relative to the original (before the convolution). This decrease in total power will result in dampening the error fluctuations, thereby reducing the amplitude of the errors relative to our original noise realization. To correct for the decrease in total angular power of the map and conserve the power of the error fluctuations, we scale the coefficients by $\sqrt{\sum_l C_{l_0} / \sum_l C_{l_{\rm smooth}}}$ where $\sum_l C_{l_{0}}$ is the total power of the error realisation before the smoothing process and $\sum_l C_{l_{\rm smooth}}$ is the total power of the map after smoothing. We then compute the inverse transform of Equation \ref{eq:alm} to return to configuration space. 
We then re-adjust the mean by adding a uniform offset of $3\,\textrm{K}$ chosen to match the zero level offset in \cite{Haslam408}. We do not expect this error realization to exactly match reality, however we use it as our fiducial error realisation to study its effect on our data analysis pipeline. Note that our framework is applicable to any error scenario. Thus the effectiveness of our framework to precisely recover the 21-cm signal as shown in Section \ref{sec:ResultsSpectrallyComplex} are not motivated by our specific choice for the noise realization.

\section{Modeling Amplitude Perturbations in Our Foreground Model}
\label{sec:AmpltiudeScaleFactorsBigSection}
Our foreground model described in Section \ref{sssec:spectral_model} relies on radio sky maps to construct our basemap. Since these radio sky maps are derived from measurements with no detailed error maps, the true amplitude of the radio sky will deviate from that of our foreground basemap by an amount that, given current uncertainties on low frequency sky maps, will (almost) certainly preclude unbiased recovery of the 21 cm signal within a forward modelling analysis of the type described in Section \ref{sssec:DataSimulations}. To account for these temperature fluctuations relative to our model, as well as any global temperature offsets in our map, we introduce a new foreground model, with amplitude correction factors (amplitude scale factors). Along with a linear offset term, we show that our model can account for these observational uncertainties. In this section we introduce our amplitude scale factor model and discuss our optimization procedure, as well as possible limitations of our framework.

\subsection{Amplitude Scale Factors}
\label{sssec:AmpltiudeScaleFactors}

In this section we introduce our amplitude scale factor model. 
Recall in Equation \ref{eq:REACHForegroundModelSpectralMasks} that the spectral map was split into $N_{\beta, j}$ regions where each region has a uniform spectral index $\beta$ which is then fit for. The spectral map used to define these regions was shown in Figure \ref{fig:BetaMap}. Example regions for $N_\beta = 4$ and $N_\beta = 8$ are shown on the left side of Figure \ref{fig:RegionsSplit}.
Our approach to account for amplitude perturbations in the basemap is analogous to this framework, but now applied to fitting for the true base-map in place of the spectral structure on the sky. To implement amplitude correction factors (hereafter amplitude scale factors) into our foreground model, we split the foreground basemap into $N_a$ subregions, with each region having an associated multiplicative scale factor $a_i$ that adjusts the temperature in the foreground basemap for that subregion. Note that the $N_a$ amplitude scale factor regions are not coincident with the $N_\beta$ spectral regions and thus the $M_{\beta,j}$ masks do not align with the $N_a$ scale factor regions. This is a result of not using the spectral map in Figure \ref{fig:BetaMap} to define the $N_a$ scale factor regions.  On the right side of Figure \ref{fig:RegionsSplit} we show example amplitude scale factor regions defined using the Haslam basemap. Note that although they appear similar since they both follow galactic morphology, they are not identical. Having non-coincident $N_a$ and $N_\beta$ regions provides our model more flexibility leading to higher fidelity fits. Our updated foreground model, including these amplitude scale factors, is given by,
\begin{dmath}
T_{\rm FG}(\theta,\phi, \nu)  = \sum_i^{N_a} \sum_j^{N_\beta} \bigl[a_i M_{a, i}(\theta,\phi) M_{\beta, j}(\theta,\phi) (T_{\rm base}(\theta, \phi, \nu_0)- T_{\rm CMB}) + \gamma_{\rm offset} \bigr](\nu/\nu_0)^{-\beta_j} 
\label{eq:FullForegroundModelNaAndBeta}
\end{dmath}
where $M_{a,i}(\theta,\phi)$ are masks applied to the foreground basemap which take on the value of 1 or 0 depending on whether a pixel is part of the $i$th scale factor region or not. All pixels in the map comprising a scale factor region are then multiplied by $a_i$. Thus any amplitude perturbations in the data relative to the foreground basemap can be adjusted for by scaling the model in that region by $a_i$. Similarly $N_\beta$ are the number of spectral regions and $M_{\beta,j}(\theta,\phi)$ are the associated masks for the spectral regions (as discussed in Section \ref{sssec:DataSimulations}), where  $\beta_j$ are the associated spectral values for that spectral subregion. Thus there are $N_\beta$ spectral parameters and $N_a$ amplitude scale factor parameters in our foreground model. Note that in the special case where the $N_\beta$ and $N_a$ regions are perfectly coincident, the $N_\beta$ and $N_a$ regions now refer to the same areas of the sky but one still has $N_\beta$ and $N_a$ parameters to adjust for in the foreground model. The term $\gamma_{\rm offset}$ is a constant term added to the foreground model to adjust for any temperature offsets in the basemap relative to the data \footnote{In principle, an temperature offset in the model basemap relative to the data can also be fit for using only the amplitude scale factors. However it is computationally simpler and conceptually more intuitive to separate these effects with the inclusion of a linear offset term $\gamma_{\rm offset}$}. Note that an error scenario consisting only of a uniform offset corresponds to the errors in the rightmost columns in the left and right panels of Figure \ref{fig:PassFail}. 

The value of the amplitude scale factor $a_i$, in each subregion is adjusted along with the offset $\gamma$, $N_\beta$ parameters and 21-cm signal parameters using the Bayesian inference framework described in Section \ref{sssec:BayesTheoremPolychord} using the likelihood in Equation \ref{eq:LikelihoodTimeAveraged}, with $T_{\rm model}$ now referring to our modified foreground model (i.e. Equation \ref{eq:FullForegroundModelNaAndBeta}). To determine the optimal number of regions $N_a$ required to fit the data, we perform the inference analysis described in Section \ref{sssec:BayesTheoremPolychord} for a range of models with different $N_a$. For each model we compute the Bayesian Evidence described in Section \ref{sssec:BayesTheoremPolychord}. The suitable number of regions to split the prior map into is then determined post analysis by selecting the value of $N_a$ and $N_\beta$ that maximizes the Bayesian Evidence. The Bayesian Evidence is penalized when additional complexity is added without substantially improving the fit \citep{Mackay92}. Determining $N_a$ and $N_\beta$ using his method also allows us to avoid using more amplitude scale factors than are required to fit the data and thus prevents an overfitting scenario. For other applications of this methodology in the field of 21-cm cosmology the reader is encouraged to see e.g. \cite{PeterPober2020, MurrayPeterEDGES, PeterPoberI, PeterPoberII,  Roque}.

\subsection{Prior Maps \& Limitations}
\label{sssec:HowToSplitRegions}
In the previous section we introduced a foreground model which has the flexibility to account for multiplicative errors in the true temperature of the foregrounds relative to our model. In this section we discuss the importance of an appropriate error prior map to define the number of regions. To define the regions on the sky we require a map which describes the spatial arrangement and amplitude of the errors. This map essentially acts a prior on the amplitude scale factors and thus we refer to the map used to define the regions as the ``error prior map''. To construct the regions we bin the pixels of the error prior map into $N_a$ segments, where $N_a$ are the number of amplitude scale factors. The boundaries of the $i$th bin is determined by computing the temperature corresponding to the 100i/$N_a$ percentile in error, where the width of each bin is 100/$N_a$. Therefore the $i$th bin corresponds to pixels in the map lying between the 100($i- 1$)/$N_a$ and 100$i$/$N_a$ percentile. The pixels in each bin are then mapped to their corresponding locations on the sky. Note that this procedure can be applied using any map and thus in principle any map can be used to define the $N_a$ sub-regions. However since the scale factors are temperature multiplicative factors, grouping regions of the sky with similar temperature perturbations increases the effectiveness of the amplitude scale factors within this framework. In Section \ref{sec:ApplicationToSpectrallUniform} we show that having $N_a$ sub-regions to coincide with the true fractional errors in the foreground map optimizes this approach by both reducing the number of regions and improving the fits. Thus using our best estimate of the foreground errors as our error prior is an important step in our procedure. We then proceed in our analysis by presenting two scenarios which act as limiting cases. In a conservative approach we use the 408MHz Haslam basemap to define the regions. In the right column of Figure \ref{fig:RegionsSplit} we show example regions using the Haslam basemap. The upper right panel corresponds to $N_a = 4$ while the lower right column corresponds to $N_a = 8$. Using this map as our prior map is conservative since the morphology of the absolute error map roughly follows the galactic structure and so represents a scenario where we haven't included any additional information in our framework regarding the errors. In our second scenario, we consider the limiting case of having perfect knowledge of the spatial structure (but not necessarily amplitude) of the errors in the sky model. For this, we use the true fractional error map to define the region defined as 
\begin{equation}
    \label{eq:fractional_error_definition}
    f(\theta, \phi) = \frac{T_{\rm base}(\theta, \phi) + \varepsilon (\theta,\phi)}{T_{\rm base}(\theta, \phi)}
\end{equation}
where $T_{\rm base}$ refers to the unperturbed basemap and $\varepsilon$ refers to error realisation contained in the basemap. Information on the per-pixel covariances in the eGSM map may be available in the future and thus would inform our models regarding the correlation structure of the errors. For now, we use this perfect prior map to illustrate the best-case scenario for the performance of this approach. Unless otherwise mentioned we define the $N_a$ scale factor regions using Haslam basemap from \cite{Haslam408}.

In the previous section we discussed how the Bayesian Evidence can be used to determine the optimal number of regions. One limitation in using this framework to define the regions is that adjacent $N_a$ models are not subsets of another, i.e. a model with $N_a $ regions is not a subset of a foreground model with $N_a + 1$ regions. As a result, increasing the complexity of the foregrounds models by increasing  $N_a$ does not build on top of another one; instead, each model is an entirely new reconfiguration of regions\footnote{Due to the percentile splitting, the $n^{2m}th$, (where $m = 0, 1, 2 ..$) regions are all subsets of one another.}. This impacts how the Bayesian Evidence of each model depends on $N_a$.  In general one would expect that as more parameters are added to the model, the fit of the model to data should improve, thereby increasing the Bayesian evidence up until the additional complexity is no longer required to describe the data, at which point the Bayesian Evidence decreases with $N_a$. Since, for most values of $N_a$ the amplitude scale factor regions are not nested, we no longer expect a strict monotonic increase of the Bayesian Evidence as a function of $N_a$. This effect will be more pronounced when the error prior map used to define the regions does not coincide with the true temperature perturbations errors in the data. In this scenarios the placement of the regions with respect to the true fractional errors is suboptimal and so there will be certain models with the scale factor regions that are, by chance, better aligned with the true errors. In contrast, in the perfect error structure information scenario the region placement is already aligned with the errors in the map and so the variation in $\ln(Z)$ from model to model is smaller. Moving forward we denote the scenario as having an ideal prior map.

\subsection{Map Reconstruction}
\label{sssec:MapReconstruction}
Outside of the context of a global 21-cm signal experiment, one may use this approach to recover a model map of the radio sky without the error perturbations, i.e. performing foreground map reconstruction. To do this one uses the mean posterior model map as the  best-fitting model of the sky as seen by the instrument.
However using this framework for map construction should be used with discretion. As discussed in Section \ref{sec:AmpltiudeScaleFactorsBigSection}, the mean temperature on the sky is the key metric of our model which is being compared to the dataset within our Bayesian inference framework.  As a result our inference framework finds the optimally fitting values of $a_i$ in each sub-region of our foreground model such that the mean temperature of the sky in our model fits the mean temperature of the simulated dataset. Therefore one finds that the optimal value of $a_i$ in the $i$th sub-region is the spatially averaged value of all the amplitude perturbations within the $i$th sub-region. However because each sub region will contain a distribution of temperature perturbations, and only the mean of $a_i$ is chosen, the boundaries between adjacent sub-regions will be disjointed. Our framework thus only adjusts the mean in each region, which does not prioritize the smoothing between boundaries of regions. Additional operations would be required to smooth out the map. However increasing the number of scale factor regions $N_a$ will create a smoother interface at the boundaries improving the model reconstruction. 

The performance of map reconstruction is also dependent on the error prior map used to define the scale factor sub-regions. Using the true fractional error map will decrease the disjointedness at the boundary between regions and increase the performance of the map reconstruction. This is because any region defined using the true fractional errors contained in the data will have a smaller range of amplitude perturbations contained in that region and therefore create a more seamless transition across the boundary. Note that disjointed boundaries between regions do not impact the performance of our framework or cause any systematics in our analysis because the temperature of each sub-region in the sky is summed according to Equation \ref{eq:FullForegroundModelNaAndBeta} without any data analysis procedure operating on the boundaries.



\section{Isolated Amplitude Errors}
\label{sec:IsolatedAmplitudeErrors}
In this section we use our foreground model to fit a dataset which was constructed using a basemap containing the error realizations derived in Section \ref{sssec:RealisticErrors}. A 21-cm signal is included in the dataset using Equation \ref{eq:21-cmGaussian} with amplitude, standard deviation and centering frequency $A_{\rm 21} = 0.155$K, $\sigma_{21} = 15$MHz, and $\nu_{21} = 85$MHz respectively, which are equivalent to the fiducial signal model in \cite{REACHSpectralModel}. Note that it has been shown by \cite{DominicInstrument} that varying the amplitude and central frequency of the fiducial signal can affect the bias in the recovered signal in the REACH pipeline. We expect similar qualitative results to hold in our amplitude scale factor framework. Since the focus of the paper is to introduce our amplitude scale factor framework and not practically explore signal recovery, we keep $A_{\rm 21}$, $\sigma_{21}$, and $\nu_{21}$ fixed throughout the analysis.  Note that we include a model of the conical log spiral beam when constructing our simulated data sets and when forward modelling the data to recover the 21-cm signal estimates (see Figures 3 and Figure 4 in \citealt{REACHSpectralModel} for the details of our beam model). For example, see Equation \ref{eq:DatasetAfterBeam} where we take the sky, rotate it to zenith multiply by the beam $D$ at each frequency channel and at every point in space and then integrate across the sky. Also in Equation \ref{eq:SkyModelAfterBeam} we include the beam in our forward model of the sky. Thus even a flat spectrum would appear chromatic in our analyses. However note that improper modelling of this antenna beam can also lead to biased recovery of the 21cm signal. This will be explored in upcoming work (Anstey et al. in prep). In this upcoming work it will be also be shown that parametrizing the beam can account for errors in the beam modelling and recover unbiased estimates of the 21cm signal.
Thus in a realistic observational pipeline one must jointly fit for the beam parameters and the $N_a$ amplitude scale factors. In this paper we assume that the antenna beam is known without error in order to isolate the effect of amplitude scale factors on the data analysis pipeline. We also make the simplifying assumption that there is no spatial dependence to the spectral index. We fix $\beta = 2.7$ in the simulated dataset as well as in the foreground model used to fit the dataset. We therefore isolate the effect that amplitude perturbations have on the analysis without the effect of spectral errors. We consider the case of spatially varying spectral index in the following sections. In Section
\ref{sec:ApplicationToSpectrallUniform5pt1} we apply our amplitude scale factor framework to a single error scenario from Section \ref{sssec:RealisticErrors} to highlight the general trends between the number of amplitude scale factor regions and the Bayesian Evidence. We compute optimal number of regions to describe the simulated dataset and show the recovered 21-cm signal for that model. In Section \ref{sec:ApplicationToSpectrallUniform} we then show how the optimal number of amplitude scale factors changes with the error properties in Section \ref{sssec:RealisticErrors}. 
In Section \ref{sssec:InformativePrior} we discuss the effect that using the true fractional error map discussed in Section \ref{sssec:HowToSplitRegions} as a prior map to define the $N_a$ regions has on the analysis.

\subsection{Spectrally Uniform}
\label{sec:ApplicationToSpectrallUniform5pt1}

 \begin{figure*}
  \includegraphics[width=0.9\textwidth]{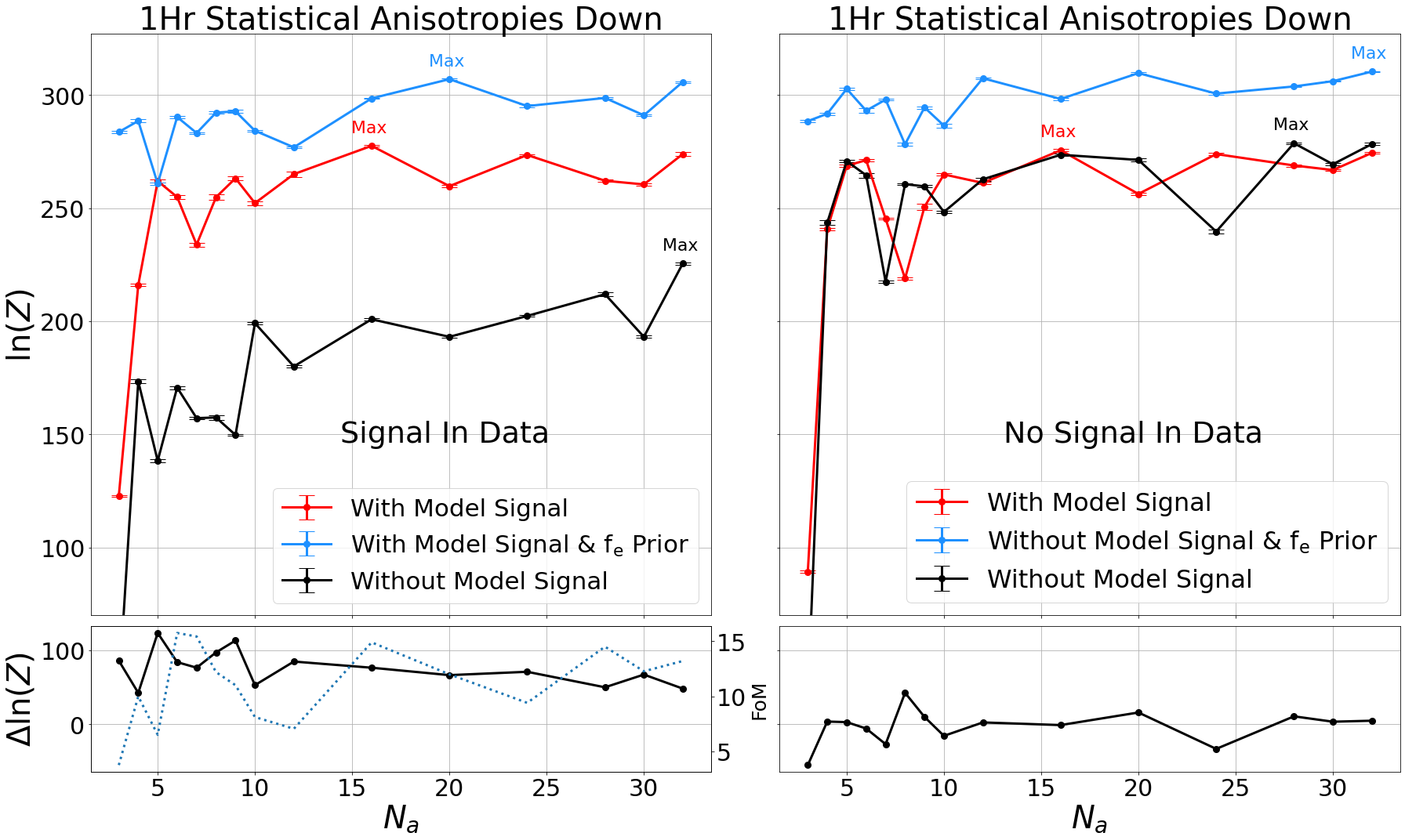}
  \caption{The Bayesian evidence, $\ln(Z)$, as a function of the number of amplitude scale factor regions $N_a$. The black curve indicates a scenario where no 21-cm signal is included into the sky model while the blue curves indicate sky models when a 21-cm signal is included. Left panel: 21-cm signal is included into the dataset. We see that a 21-cm signal model component is preferred since the maximum value of $\ln(Z)$ originated from the red curve. The $\Delta \ln(Z)$ curve is the difference in Bayesian evidence between the blue curve and black curve. The dotted blue curve indicates the FoM (see Equation \ref{eq:FoM}) for the recovered signal model (red curve). A larger FoM suggests a more accurately recovered signal. Right panel: no 21-cm signal is included in the dataset, in which case a sky model without a 21-cm signal component is preferred since the maximum value of $\ln(Z)$ occurs in the black curve. The blue curves correspond to a model where an ideal prior map (i.e. the fractional errors in Equation \ref{eq:fractional_error_definition}) are used to define the regions. The $\Delta \ln(Z)$ curve is the difference in Bayesian evidence between the black curve and red curve.  We use a one hour observation duration in these simulations. Note the area of high foreground map uncertainty in the lower right portion of the basemap in the upper left of Figure \ref{fig:FiducialErrorMap} (this area is present in each map, but is most pronounced in the unsmoothed map). This area of high foreground map uncertainity is due to many low frequency surveys missing the southern pole which influences the eGSM model. We choose the one hour observation to occur at a time where this area is mostly below the horizon. Since regions below the horizon are removed during our simulated observation, this area of high uncertainty in the lower right of the eGSM basemap does not enter our analysis. In particular, we avoid this part of the map since regions with least foreground uncertainty are most promising for 21-cm signal recovery (this reduces the potential for systematics and bias in the analysis which also reduces the foreground model complexity required to model the 21-cm signal to high fidelity).}
  \label{fig:SpectrallyUniformWSignalWOSignal}
\end{figure*}

 \begin{figure*}
  \includegraphics[width=0.95\textwidth]{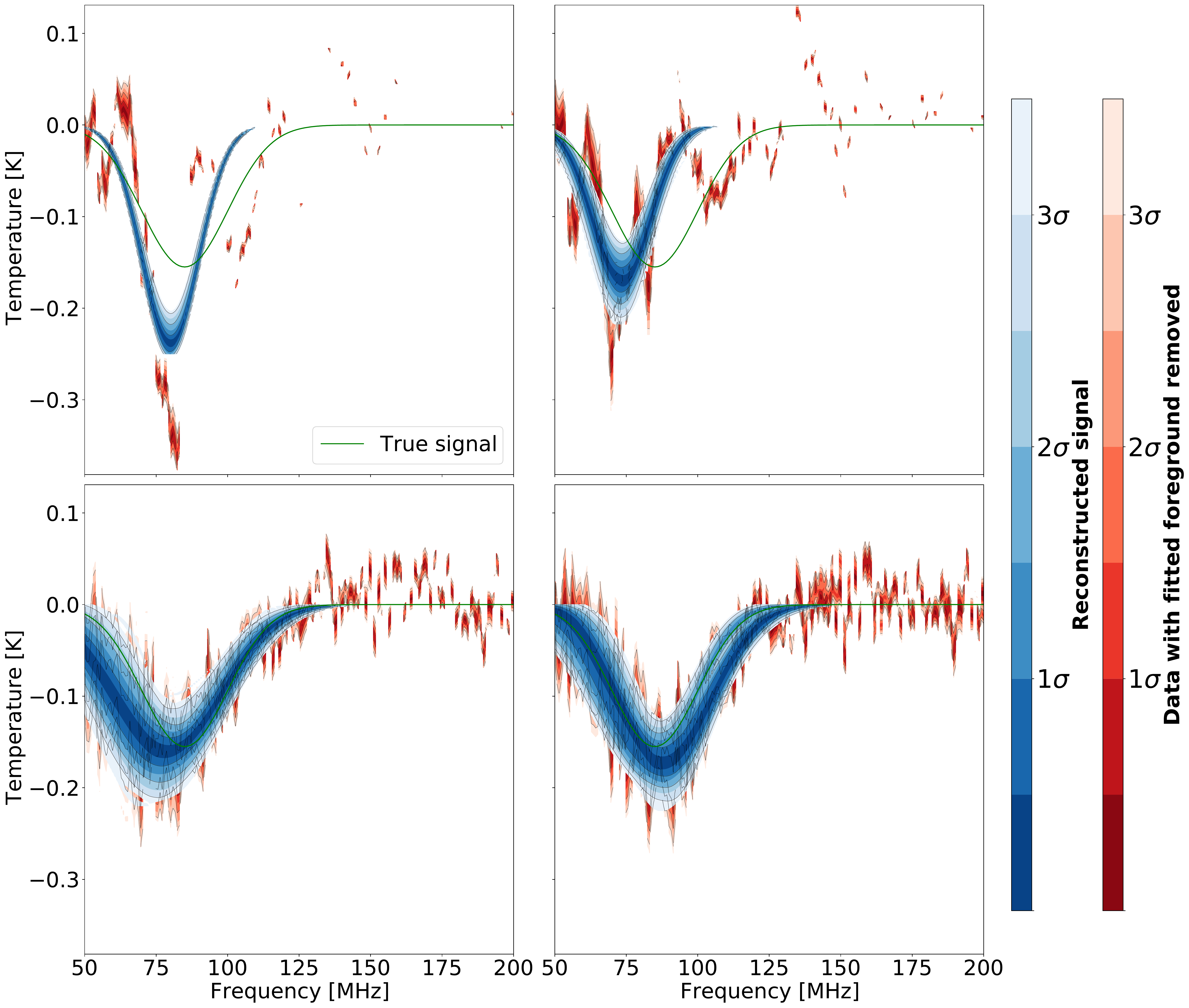}
  \caption{Each plot shows the results of fitting our foreground model to the simulated data discussed in Section \ref{sec:ApplicationToSpectrallUniform5pt1}. The upper left panel corresponds to $N_a = 1$. Moving clockwise, we have models  $N_a = 3 , 16, 32$. The blue curves correspond to the fitted 21-cm signal while the red curves correspond to the residuals of the dataset after the fitted foregrounds are removed.   }
  \label{fig:ReconstructedSignals}
\end{figure*}

 \begin{figure*}
  \includegraphics[width=0.9\textwidth]{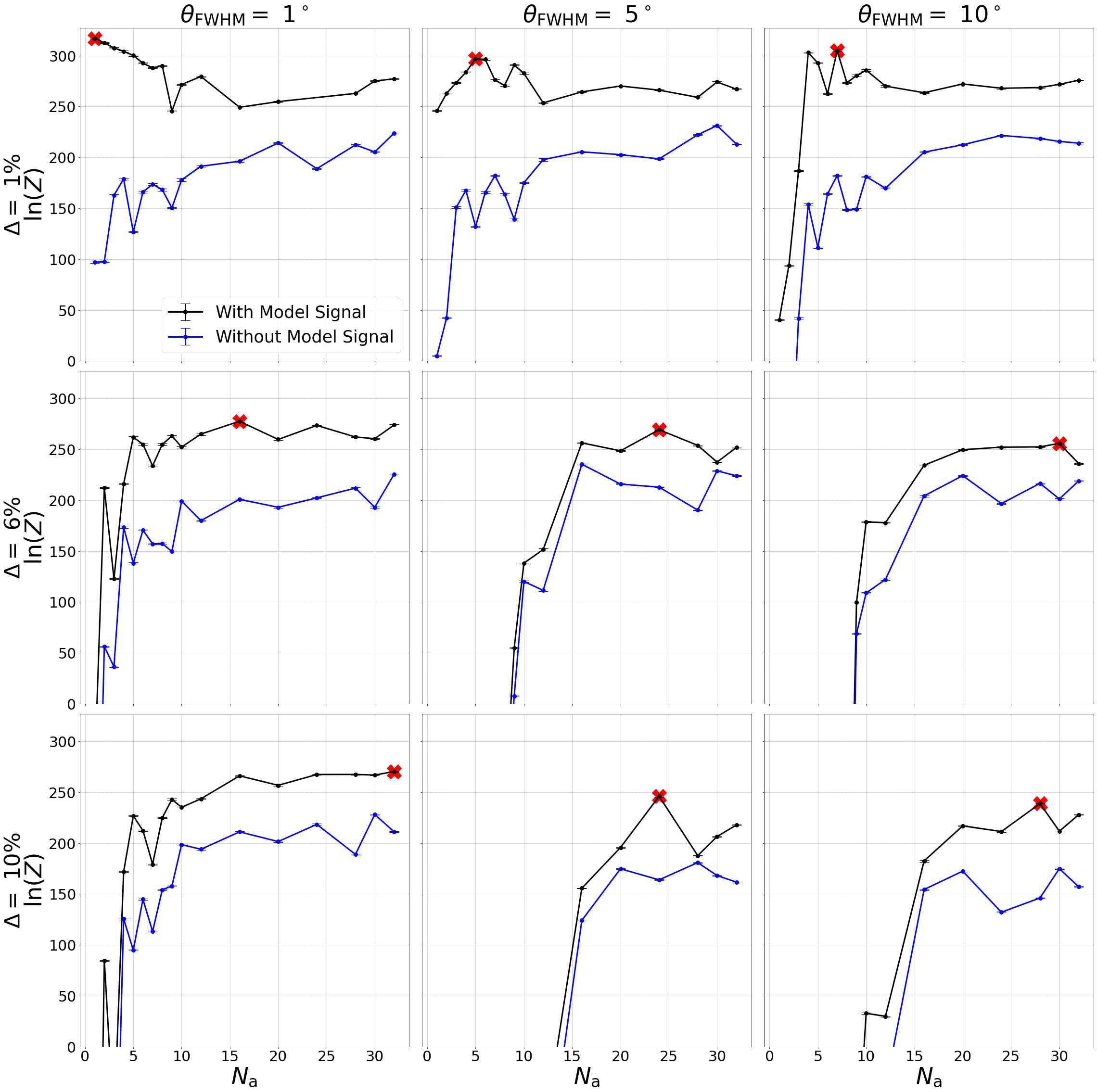}
  \caption{The Bayesian evidence, $\ln(Z)$, as a function of the number of amplitude scale factor regions $N_a$ for each error scenario described in Section \ref{sssec:RealisticErrors}. The black curves indicate a scenario where a 21-cm signal is included into the sky model while the blue curves indicate sky models where no 21-cm signal is included. Note that 21-cm signal is included into the dataset leading to black curves having larger $\ln(Z)$ than the blue curves. We observe that the number of amplitude scale factor regions required to maximize $\ln(Z)$ tends to increase with error fluctuation amplitude $\Delta$ and correlation length $\theta_{\rm FWHM}$. }
  \label{fig:ParametrizedErrorsFlucSmooth}
\end{figure*}

 \begin{figure}
  \includegraphics[width=0.5\textwidth]{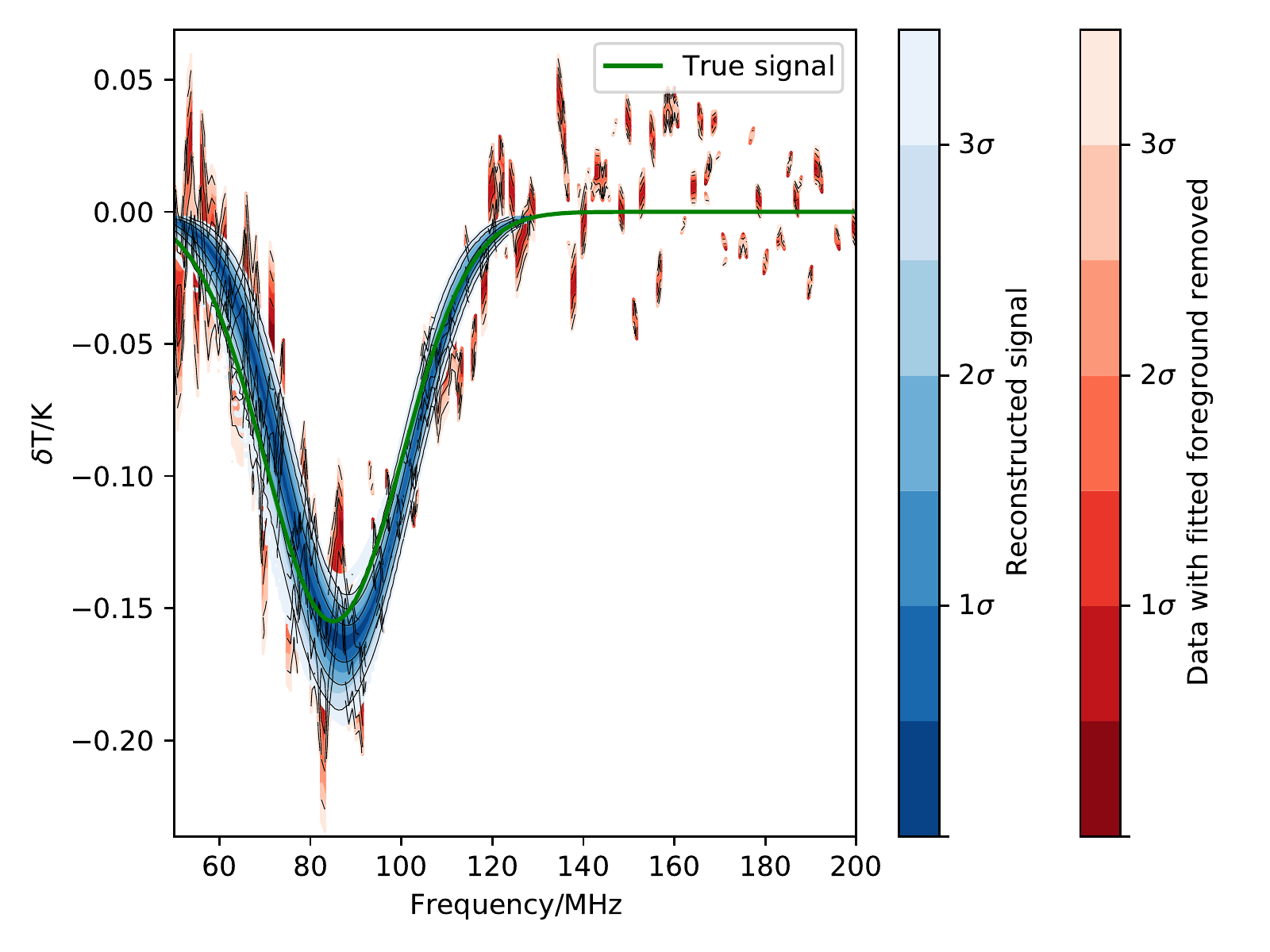}
  \caption{The reconstructed signal (blue) and residuals (red) for the maximum Bayesian evidence model, $N_a = 20$ corresponding to the blue curve on the left panel of Figure \ref{fig:SpectrallyUniformWSignalWOSignal}. In this model the regions were defined using an ideal prior, i.e. the fractional errors in the basemap as defined by Equation \ref{eq:fractional_error_definition}. Using this prior represents a best case scenario where the true fractional errors in the amplitude of the basemap are known. Note that the residuals are smaller, and more consistent with the true signal (green) as compared to Figure \ref{fig:ReconstructedSignals}.  }
  \label{fig:InfResiduals}
\end{figure}


In the previous section we introduced our amplitude scale factor framework. In this section we select one error scenario from Section \ref{sssec:RealisticErrors} to highlight our model and illustrate the general trends of the number of scale factors $N_a$ and the Bayesian evidence. In the following section we apply our framework to all error scenarios introduced in Section \ref{sssec:RealisticErrors}. In Figure \ref{fig:SpectrallyUniformWSignalWOSignal} we show the Bayesian evidence as a function of $N_a$ for $\Delta = 6\%$ and $\theta_{\rm FWHM} = 1^\circ$. On the left panel 21-cm signal is included in the dataset while on the right panel no 21-cm signal is included in the dataset. For each case, the dataset is fit with a foreground model that has $N_a$ amplitude scale factors, one offset parameter $\gamma$ and either a signal model (red curves) or no signal model (black curve). Starting from $N_a = 1$ in the left panel, we can see that as more degrees of freedom are added to the model, $\ln(Z)$ increases rapidly until where we see that $N_a = 16$ maximizes the Bayesian Evidence. The residuals for $N_a = 1$ and $N_a = 16$ are shown in the first column of Figure \ref{fig:ReconstructedSignals}. From this figure, we can see that the systematic created by the amplitude perturbations is substantially reduced as we increase the number of scale factors. Since one scale factor is equivalent to adjusting the mean of the foreground map, this result implies that the presence of the spatial fluctuations of the foreground errors about the mean of the error map also contribute to the bias in the signal recovery within the analysis. Our amplitude scale factor framework can accommodate these fluctuations with more precision by adding more regions to the foreground model. Thus the size of the regions shrink and can begin to accommodate smaller features in the error map. In Figure \ref{fig:ReconstructedSignals} we show the residuals for the $N_a = 3$ model (first row, second column) and see that the bias in the signal recovery has shrunk relative to the $N_a = 1$ model (first row, first column). In the lower left of Figure \ref{fig:ReconstructedSignals} we show the residuals for the model which maximizes the Bayesian evidence ($N_a = 16$). In this case the amplitude, central frequency and standard deviation of the reconstructed signal are consistent with the true values at $1\sigma$. Referring back to Figure \ref{fig:ParametrizedErrorsFlucSmooth} we see that as we continue to increase the regions beyond $N_a = 16$ the Bayesian Evidence slowly decreases. In the lower right panel of Figure \ref{fig:ReconstructedSignals} we show the residuals for the $N_a = 32$. The amplitude and standard deviation of the systematics have decreased further; however, the additional flexibility added to the system beyond $N_a = 16$ is not required to fit the data, and as a result there is a statistical cost to adding more parameters. This is due to to Equation \ref{eq:BruteForceEvidence}, whereby we can see that adding additional parameters increases the prior volume which penalizes the Bayesian evidence. Note that as discussed in Section \ref{sssec:other_error_scenarios} observation times when the galaxy is above the horizon produce larger systematics in the data as compared to when the galaxy is below the horizon. We observe that configurations when the galaxy is above the horizon translates to requiring more regions $N_a$ to maximize the Bayesian evidence as compared to galaxy down configurations. In each case our model is applicable however one should expect that the $N_{\rm a}$ required to maximize the Bayesian Evidence will depend on observation times. Also note the scale difference of the red and black curves in the left and right panels in Figure \ref{fig:SpectrallyUniformWSignalWOSignal}: a Gaussian 21-cm model component can be used as an extra degree of freedom to fit foreground systematics (see Figure \ref{fig:SignalRecoveryMaxEvidenceSpecComplex}).  However a foreground model does not have the ability to fit a 21-cm Gaussian signal. Thus 
\begin{enumerate}
    \item if there is a 21-cm signal in the data, we will notice a large difference in evidence between sky models which include/do not include 21-cm components .
    \item if there is no 21-cm signal in the data we will notice smaller differences in evidence between sky models which include/do not include 21-cm components. 
\end{enumerate} 
Note that simulations with a more flexible 21-cm model might therefore covary with systematics making it more difficult to separate each component in the modeling. We plan to investigate this further in future work.

To quantify the fit of the recovered signal to the true signal we compute the Figure of Merit (FoM) introduced in \cite{REACHSpectralModel} and defined as 
\begin{equation}
    \label{eq:FoM}
    \textrm{FoM} = \frac{A_{\rm amp}}{ \sqrt{ \frac{1}{N_{\nu}} \sum_{\nu} \left ( T_{\rm true} - T_{\rm fit} \right )^2   } }
\end{equation}
where $N_{\rm \nu} = 151$ are the number of frequencies, $A_{\rm amp} = 0.155K$ is the fiducial amplitude of the signal and $T_{\rm true}$ and $T_{\rm fit}$ are the true signal and recovered signal respectively. A larger FoM suggests a more accurately recovered 21cm signal. In the bottom panels of Figure \ref{fig:SpectrallyUniformWSignalWOSignal} we show the FoM as a function of $N_a$ for our models. 
Note that because the regions between two consecutive models $N_a$ and $N_a + 1$ are entirely reconfigured, sky models with $N_a$ regions are not a subset of those used in an analysis with $N_a+1$ regions. Thus there is no requirement that the FoM or the Bayesian Evidence $\ln(Z)$ must monotonically decrease (or increase) with $N_a$. Therefore fluctuations in the FoM and $\ln(Z)$ after the Bayesian Evidence maximizing model are expected.

\subsection{Effect of the Amplitude of Fluctuations $\Delta$ and Correlation Length $\theta_{\rm FWHM}$ of the Errors on the Analysis}
\label{sec:ApplicationToSpectrallUniform}

In Section \ref{sssec:RealisticErrors} we introduced our fiducial foreground error model, which was parametrized in terms of the amplitude of the fluctuations $\Delta$, and the FWHM of correlated features $\theta_{\rm FWHM}$. In this Section we apply our amplitude scale factor foreground model described in section \ref{sssec:AmpltiudeScaleFactors} to a dataset constructed using a foreground map which contains these fiducial error scenarios. In Figure \ref{fig:ParametrizedErrorsFlucSmooth} we show the Bayesian Evidence for each scenario.  
We see that the number of amplitude scale factor regions required to maximize the Bayesian Evidence depends on the level of the fluctuations $\Delta$ and $\theta_{\rm FWHM}$ of the error realisation in the datasets. From Figure  \ref{fig:ParametrizedErrorsFlucSmooth} it is evident that for fixed $\theta_{\rm FWHM}$, increasing $\Delta$ of the errors in the dataset increases the optimal number of amplitude scale factors required to fit the data. This is an expected result. Consider a foreground model with $N_a$ scale factors regions. Any systematics remaining in the analysis are due to small fractional differences in temperature of the foreground dataset relative to the foreground model. Thus uniformly increasing the mean fractional differences of the entire foreground map used to construct the dataset relative to the basemap will scale any systematics remaining in the analysis. Thus scaling the temperature fluctuations will require more precision in the foreground model to reduce the systematic. This trend is independent of how the scale factor regions are defined. This behaviour is independent of the prior map used to define the regions.

It is also clear from Figure \ref{fig:ParametrizedErrorsFlucSmooth} that increasing the FWHM of the error structures results in requiring more scale factors to maximize the Bayesian evidence. To understand this, consider a region of the sky containing many small correlation structures such as in Figure \ref{fig:FiducialErrorMap}. This leads to a scenario where many independent error realisations can be contained within a region and thus tend to average out, i.e. the noise is driven to Gaussian random noise. However as the FWHM of the correlation structures increases, a single error feature may overlap into multiple regions, thus requiring more regions to isolate and localize independent error realisation features. 

Note that we do not consider multiple values of global temperature offsets in the foreground errors since the linear offset term $\gamma$ in Equation \ref{eq:FullForegroundModelNaAndBeta} allows our model to be robust to these scenarios. Thus our model allows the analysis of an error scenario with offset $3$K to be equivalent to one with $0$K offset. 

\subsection{Limiting Cases}
\label{sssec:InformativePrior}

In the previous sections, we used the 408MHz Haslam basemap to define the scale factor regions. The regions thus followed the Galactic morphology, which do not have a similar structure to true fractional errors in the basemap. In this section we illustrate the effect that using a suitable prior map to define the $N_a$ regions has on the performance on our framework. We compute the fractional error map as defined by Equation \ref{eq:fractional_error_definition} to define the $N_a$ regions. These two scenarios effectively bracket extreme scenarios, i.e. a scenario where we know precisely the morphology of the amplitude perturbations and a scenario where we simply assume the errors follow the Galactic morphology. In the future information on the per-pixel covariances in the eGSM map may be available. For now, we use this perfect prior map to illustrate the best-case scenario for the performance of this approach.

In Figure \ref{fig:SpectrallyUniformWSignalWOSignal} (blue curve), we show the evolution of the Bayesian evidence as a function of $N_a$ using regions derived from the ideal prior map for the error scenario $\Delta = 6\%$ and $\theta_{\rm FWHM} = 1^\circ$.  The behavior of the Bayesian evidence as a function of $N_a$ using the ideal prior to define the regions produce similar qualitative trends with two quantitative differences:
\begin{enumerate}
    \item The Bayesian Evidence for the maximum evidence model is larger when using the fractional error map to define the regions
    \item more regions are required for models using the fractional errors.
\end{enumerate}
These quantitative differences are due to the morphological differences between the absolute error map (which follows the galactic morphology) and the actual amplitude perturbations in the dataset. Since the fractional errors are better aligned with the true multiplicative errors in the map, the number of regions required to optimize the Bayesian Evidence is increased from $N_a = 16$ using the absolute error map to define the regions to $N_a = 20$ using the fractional errors to define the regions. The maximum Bayesian Evidence is also larger using the fractional errors to define the regions.  


We show the residuals for $N_a = 20$ maximum evidence model in Figure \ref{fig:InfResiduals}. Referring to Figure \ref{fig:InfResiduals} we see that the residuals are smaller and more ``noise-like'' using the fractional error map to define the regions as compared to using galactic morphology to define the errors. When using the galactic morphology to define the regions,  the amplitude scale factor regions do not line up properly with the true fractional errors in the map. As the result, each region defined using the Haslam basemap encompasses a larger distribution of fractional errors. As a result each region cannot capture the finer error features in the foreground error map. This also results in more regions being required to describe the error structure present. Therefore we continue to see a more substantive increase in the residuals between the reconstructed 21-cm signal and the true signal as we go to large $N_a$ compared to what we saw for large $N_a$ models in the factional error prior (where the benefit to the fits were smaller). 

The blue curve on the right panel of Figure \ref{fig:SpectrallyUniformWSignalWOSignal} shows the Bayesian evidence for a sky model without a 21-cm model component and using the fractional errors to define the regions. Here we can again see that this also results in increasing the number of scale factor regions required to maximize the Bayesian evidence from $N_a = 28$ (black curve) to $N_a = 32$ (blue curve).

\section{Spatially Varying Spectral Index}
\label{sec:ResultsSpectrallyComplex}

 \begin{figure*}
  \includegraphics[width=0.9\textwidth]{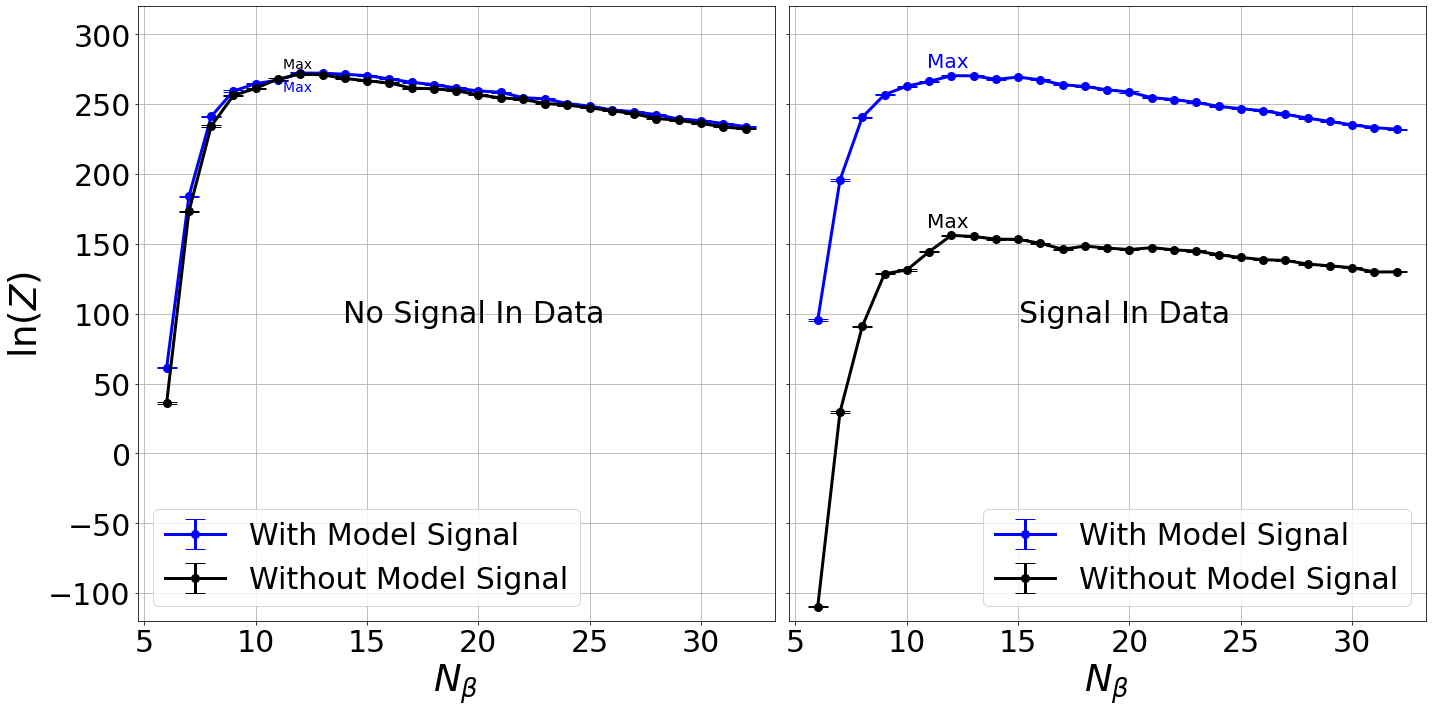}
  \caption{The Bayesian evidence as a function of the number of spectral parameters $N_{\beta}$. The black curves indicate scenarios where a 21-cm signal is included into the sky model while the blue curves indicate sky models where no 21-cm signal is included. Left panel: no 21-cm signal is included into the dataset. We see that a model without a 21-cm signal model component is preferred since the maximum value of $\ln(Z)$ originates from the black curve. Quantitatively, the the maximum value of $\ln(Z)$  for the black curve is $3$ log units larger than the blue curve}. Right panel: a 21-cm signal is included in the dataset, in which case a sky model with a 21-cm signal component is preferred since the maximum value of $\ln(Z)$ occurs in the blue curve.
  \label{fig:LogZNbNoAmplitudeErrors}
\end{figure*}

 \begin{figure*}
  \includegraphics[width=0.85\textwidth]{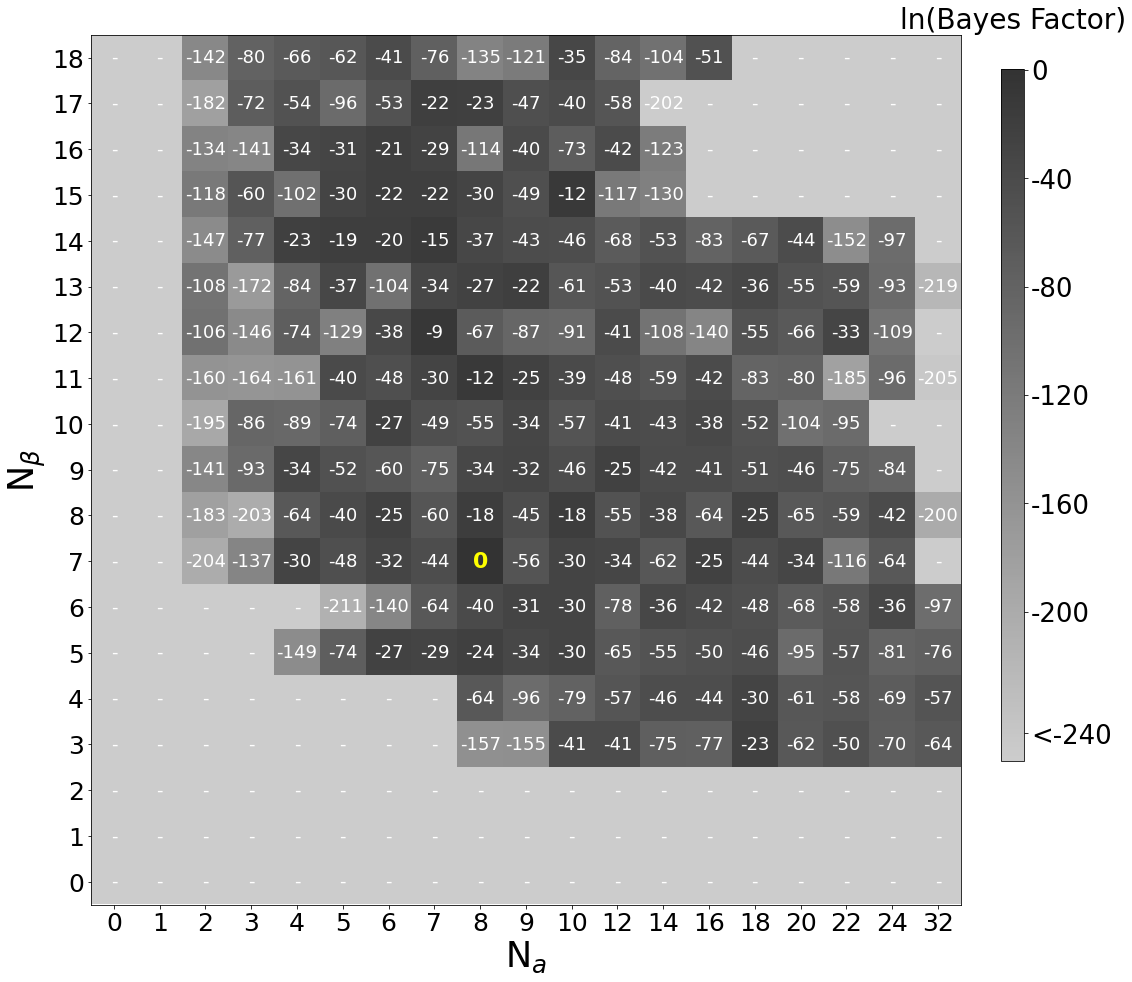}
  \caption{The logarithm of the Bayes' Factor as a function of the number of spectral parameters $N_{\beta}$ and amplitude scale factors $N_a$. The gray shading in each square indicates the value of Bayes' factor, i.e. $\ln(Z)_{\rm max} - \ln(Z)$ where the $\ln(Z)_{\rm max} = 265$ is the model $N_a = 8$ and $N_{\beta} = 7$ which maximises the Bayesian evidence. All models with Bayes' factor $< 240$ or negative Bayesian evidence are saturated at $240$. Notice that models with $N_a = 1$ are disfavoured even with large values of $N_{\rm beta}$ and vice-versa. Thus multiple spectral and scale factor regions are required. The model which maximizes the Bayesian evidence is highlighted in yellow.}
  \label{fig:Grid}
\end{figure*}
In this section we discuss our more realistic observation scenario, where the foregrounds in the simulated dataset have a spatially varying spectral index and foreground basemap errors. To simulate a realistic data analysis where we do not have access to the true fractional error, we use the 408MHz Haslam basemap to fit our simulated data. Thus we add the formalism introduced in \cite{REACHSpectralModel} which was described in Section \ref{sssec:spectral_model} to our analysis. That is, we split the spectral map in Figure \ref{fig:BetaMap} into $N_\beta$ sub regions each with uniform $\beta$ in accordance with Equation \ref{eq:FullForegroundModelNaAndBeta}.  Since the spectral regions do not need to coincide with the $N_a$ amplitude scale factor regions, there are $N_\beta \times N_a$ + 1 parameters required to be optimized in the foreground model. We again use \textsc{PolyChord} to compute the Bayesian Evidence for each model. In the interest of reducing computational cost, we use only one error scenario from Section \ref{sec:ApplicationToSpectrallUniform}. In Section \ref{ssec:PickOneErrorScenario} we select our error realisation and in Section \ref{ssec:SpecComplex} we show our results.

\subsection{Error Selection}
\label{ssec:PickOneErrorScenario}
In this section we select the most realistic error scenario from Section \ref{sssec:RealisticErrors}. To do so, note that we expect that the correlation length of the errors in the basemap will be inversely proportional to its resolution. Since in this work we use the Haslam map to construct our foreground model, we select the correlation length of our fiducial error scenario to match the $1^\circ$ FWHM resolution of the Haslam map. We therefore use the $\Delta = 6\%$, $\theta_{\rm max} = 1^\circ$ scenario from Section \ref{sec:ApplicationToSpectrallUniform}. 

Furthermore, since the resolution of the basemap is related to its base frequency, maps at lower frequency will have lower spatial resolution. For example the 150MHz empirically derived map introduced in \cite{150MHz_AllSky} has a $5^\circ$ FWHM resolution compared to $1^\circ$ for Haslam at 408MHz. Under this assumption, we would expect more correlated errors in the Landecker map. For a fixed level of power of errors in the map, we would expect this increased correlation to introduce more significant foreground systematics and, correspondingly, larger biases in the recovered 21-cm signal, in the absence of error modelling. In the context of the sky map error fitting formalism we present in this paper, we would expect a more complex error model to be required to explain the data when using the Landecker map as our base-map in an analysis of instrumental data.\footnote{This assumes that the error structure follows the Gaussian generated errors discussed in Section \ref{sssec:other_error_scenarios}}.  Recall from Section \ref{sssec:other_error_scenarios} that increasing the FWHM of the errors results in requiring more amplitude scale factor regions to maximize the Bayesian Evidence. Provided there is a negligible change to the morphology of foregrounds as a function of frequency, one can strategically use higher frequency basemaps in the foreground model to reduce the systematic impact of foreground amplitude errors in the analysis. However, increasing the base frequency of the foreground basemap might also increase the number of spectral regions $N_\beta$ required to model the spatial variation of the spectral index. To see why this is recall that the REACH observational frequencies range from $50$MHz to $170$MHz. Using a basemap at higher frequencies, i.e. Haslam at $408$MHz requires more spectral precision in the foreground model since the spectral model is extrapolated over a larger frequency range. Thus more spectral regions $N_{\beta}$ are required relative to using a basemap set at lower frequencies.

\subsection{Correlations Between $N_a$ and $N_\beta$}
\label{ssec:SpecComplex}
The datasets we consider in this section contains spectral complexity as well as errors in the amplitude component of the basemap. Thus to select the model which maximizes the Bayesian evidence, we must optimize $N_a$ and $N_\beta$ in a two dimensional grid of models. From Section \ref{sec:ApplicationToSpectrallUniform5pt1} we know that without spectral errors, $N_a = 16$ maximizes the Bayesian evidence. To gain intuition regarding how many spectral parameters might be required to sufficiently describe the spectral component of the dataset without the presence of amplitude scale factors in the basemap, we can perform the same analysis shown in Figure 10 of \cite{REACHSpectralModel}. Note that the analysis procedures in our work has two modifications as compared to the analysis done in \cite{REACHSpectralModel}. The primary difference is that we use the percentile splitting methodology discussed in Section \ref{sssec:HowToSplitRegions} to define the spectral regions. Secondly we construct our foreground model using the Haslam basemap (at 408MHz) rather than GSM (at 230MHz). For completeness we perform the analysis for datasets which contain, and do not contain, a 21-cm signal. In Figure \ref{fig:LogZNbNoAmplitudeErrors} we show the the Bayesian evidence as as a function of the number of spectral parameters $N_\beta$ when the sky signal (i.e. the simulated dataset) does not contain a 21-cm signal (left) or contains a 21-cm signal (right). The red curves correspond to a sky model that contains a 21-cm signal component while a black curve does not. Note that in the left panel, the highest evidence models occur in the black curve while on the right panel the red curves correspond to models with higher Bayesian evidence. In the left panel (without 21-cm signal in the dataset) we see that the model $N_\beta = 10$ maximizes the Bayesian evidence while on the right panel (with 21-cm signal in the dataset) we see that model $N_\beta = 12$ maximizes the Bayesian evidence. 

The simulation we consider in this section has spectral errors in addition to basemap errors. To get a sense as to where the Bayesian Evidence maximizing model might reside on a two dimensional grid of $N_\beta$ and $N_a$ models, recall that in Figure \ref{fig:SpectrallyUniformWSignalWOSignal} we found that $N_a = 16$ regions were such that they maximized the Bayesian evidence without the presence of spectral complexity in the dataset. Similarly in Figure \ref{fig:LogZNbNoAmplitudeErrors} we see that the model $N_\beta = 12$ maximizes the Bayesian evidence when the dataset does not contain amplitude errors. Thus we would expect that $N_a \le 16$ and $N_\beta \le 12$ might maximize the Bayesian evidence. The ``$\le$'' is used instead of ``$=$'' since we allow for the possibility that $N_a$ and $N_{\beta}$ might have some correlation.

\begin{figure*}
  \includegraphics[width=0.9\textwidth]{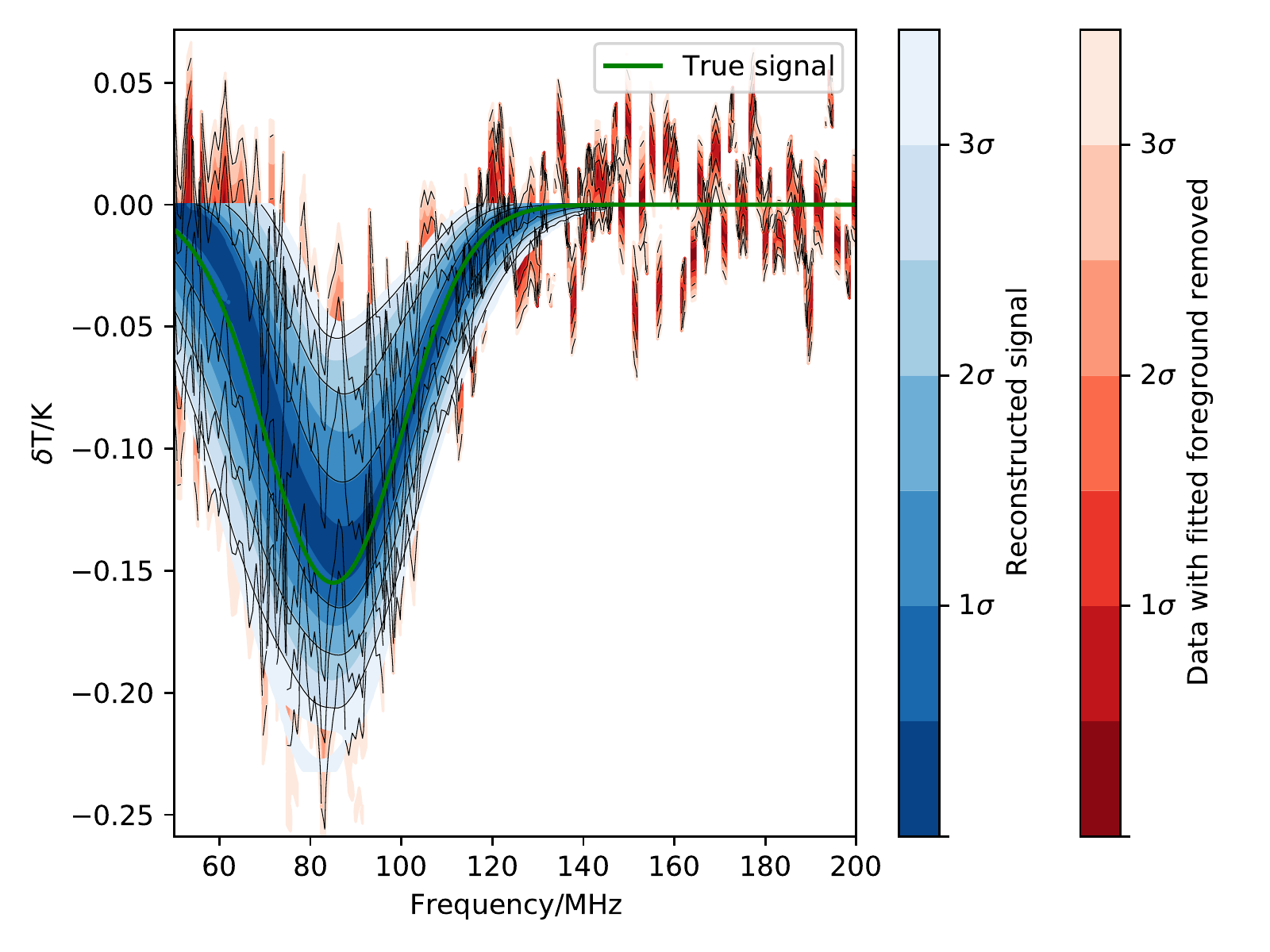}
  \caption{Signal recovery plot for the Bayesian evidence maximizing model ($N_a = 7$, $N_{\beta} = 8$) of Figure \ref{fig:Grid}. The blue curve indicates the temperature of the fitted 21-cm model while the red curve represents the dataset ($T_{\rm data}$ in Equation \ref{eq:DataSkyTemperatureFinal}) with the best fitting foreground model (i.e. Equation \ref{eq:FullForegroundModelNaAndBeta}) removed. The green curve indicates the simulated 21-cm signal that was inside the dataset.  }
  \label{fig:SignalRecoveryMaxEvidenceSpecComplex}
\end{figure*}

In Figure \ref{fig:Grid} we show the Bayes' factor, i.e. $\mathcal{B} \equiv \ln(Z_{\rm max}) - \ln(Z)$ as a function of the number of amplitude scale factor regions, $N_a$, and spectral regions $N_\beta$ used to construct the foreground model. The darkest regions of the grid correspond to models with the largest Bayesian Evidence and thus smallest Bayes' factor $\mathcal{B}$. In each square we denote the value of Bayes factor $\mathcal{B}$. Qualitatively we see that the Bayesian evidence is highest along a diagonal strip of models affirming that there is a degree of correlation between models with $N_a$ and $N_{\rm beta}$. Model $N_a = 8, N_\beta = 7$ maximizes the Bayesian evidence, and is indicated with a $0$ on its grid-point square. If multiple models have comparable Bayesian evidences, then a model averaging technique, where a weighted average is formed using their evidences should be employed. In terms of Figure \ref{fig:Grid}, the model with the next highest evidence is model $N_a = 7, N_\beta = 12$ which is $\ln(Z) = 9$ below the highest evidence. This corresponds to $9$ log units disfavouring from $N_a = 8, N_\beta = 7$ making any model averaging technique equivalent to selecting $N_a = 8, N_\beta = 7$. In Figure \ref{fig:Grid}, models with $\ln(Z) < 0$ are labeled with -. Models with this level of Bayesian evidence have systematics multiple orders of magnitude larger than the signal, making recovery of the 21-cm impossible. Note that models with $N_a = 0$ correspond to foreground models without any amplitude scale factors. Referring to the models with $N_a = 0$, i.e. first column in Figure  \ref{fig:Grid}, it is evident that increasing the number of spectral regions, even to $N_\beta = 18$ cannot compensate for the systematics caused by the foreground amplitude errors. Thus amplitude errors in the foreground basemap cannot account for using a foreground model with spectral and linear offset parameters. Similarly by examining the bottom row in Figure \ref{fig:SignalRecoveryMaxEvidenceSpecComplex} (i.e, $N_{\beta} = 1$) we can see that increasing $N_a$ to 32 regions does not account for the spectral errors in the dataset. Qualitatively, we find that above a threshold value of $N_a$ and $N_\beta$ that there is a trade-off between the number of spectral regions and number of amplitude scale factor regions.

In Figure \ref{fig:SignalRecoveryMaxEvidenceSpecComplex} we show the recovered 21-cm signal for $N_a = 8, N_\beta = 7$ which maximizes the Bayesian evidence. Recall from Section \ref{sssec:spectral_model} that our fiducial signal model has amplitude $A_{21} = 0.155K$, standard deviation $\sigma_{21} = 15$MHz and centering frequency $\nu_{21} = 85$. Note that the mean amplitude, standard deviation and centering frequency of the recovered 21-cm signal is $0.136$K, $16.1$MHz, and $85.6$MHz. Thus we find that all of the signal parameters are within 1$\sigma$ of the true values. 



\section{Summary \& Conclusion}
\label{sec:Conclusion}

Detection of the global 21-cm signal is key to our understanding of CD and the EoR. Unfortunately, foreground contaminants are orders of magnitude brighter than all theoretical models of the cosmological signal. While there do exist scientifically interesting constraints that can be made even in the face of imperfect foreground removal (especially when combined with other probes; \cite{Joelle}), realizing the full potential of global 21-cm signal measurements requires a robust foreground mitigation program. Global 21-cm signal experiments which forward model the instrument, the systematics and radio contaminants have the potential to provide us the strongest constraints on the global 21-cm signal. However forward models of the foregrounds rely on observations of the radio sky which lack detailed uncertainty characterisation. Thus approaches that forward model the foregrounds are potentially limited by our empirical knowledge of the temperature of the radio sky. In this paper we have introduced a framework which is able to account for temperature deviations in the radio sky relative to our model foreground map in global 21-cm experiments. We have constructed a foreground model where the sky is segmented into different regions, each with an associated multiplicative scale factor which can adjust the amplitude of the foregrounds for that sub-region of the sky. By fitting for the these amplitude scale factors within our Bayesian framework, we can account for temperature perturbations in the true radio sky relative to our foreground model. We select the number of sub-regions in our  model by computing the Bayesian Evidence for a large range of $N_a$ and select the model which maximizes the Bayesian evidence. Though we use the REACH experiment as our fiducial 21-cm experiment, our method is applicable to any global 21-cm experiment. 

In this paper we used a 1hr in LST observation with the log conical antenna. While we include a model of the conical log spiral beam into our dataset and forward model (see Equations \ref{eq:SkyModelAfterBeam} and \ref{eq:DatasetAfterBeam} where we assumed the beam of the conical antenna is exactly known). We also assume that the mean 21-cm signal is Gaussian in $\nu$. Each of these simplifying assumptions will impact the performance of our foreground model insofar as  
REACH will observe over a larger LST range with a dipole + log conical antenna \citep{REACH, DominicLSTDependentPipeline} where the true beam may deviate from the modeled beam. Thus, an investigation of the impact of LST-range, beam parameters, realistic 21-cm models, antenna choice and joint analysis with multiple antennas on our posterior inference (which encompasses foreground sky and global 21-cm signal) provide interesting direction we intend to pursue in future work.

Since there are no definitive models that describe the amplitude and morphology of the errors in radio sky maps, we parametrically produce a range of simulated foreground errors by drawing a Gaussian noise realisation from the 408MHz absolute error map and then parametrically smoothing the map with a Gaussian beam with various FWHM. We perform our analysis with and without spectral complexity in the sky. We find in either scenario that our framework is effective at reducing the systematics in the analysis due to the presence of the foreground map errors allowing us to recover the 21-cm signal without bias. We find that our approach is limited by our knowledge of the nature of the errors in the radio sky. Thus our work shows that more work on modelling the foreground errors is needed in order to maximize the effectiveness of our approach. Going forward we might have access to full error covariance in which case we can improve on this approach. With information regarding the correlation structure of the foreground errors we can more effectively construct the regions to match the error structure in the dataset. Since the scale factors are essentially multiplicative factors in the map, defining the regions using a map which informs the model about the location of the amplitude perturbations is an ideal scenario. Without any existing work on foreground error maps, we use the morphology of the Galaxy to define our regions. This conservative approach demonstrates how well our model can do without detailed knowledge of the true error structure. We show that even when using a conservative error map to define these regions, our method is able to construct a sufficiently high fidelity model for the foregrounds for unbiased recovery of the global 21-cm signal. Thus, we have shown that the base-map error fitting framework presented here, in combination with the spectral structure fitting methodology presented in \cite{REACHSpectralModel}, represents a powerful tool for detecting the 21cm global signal. 
 

\section*{Acknowledgements}

The authors wish to recognize the contributions of Avery Kim and are delighted to acknowledge helpful discussions with Jordan Mirocha, Hannah Fronenberg, Jonathan Sievers.  We acknowledge support from the New Frontiers in Research Fund Exploration grant program, a Natural Sciences and Engineering Research Council of Canada (NSERC) Discovery Grant and a Discovery Launch Supplement, the Sloan Research Fellowship, the William Dawson Scholarship at McGill, as well as the Canadian Institute for Advanced Research (CIFAR) Azrieli Global Scholars program. PHS was supported in part by a McGill Space Institute fellowship and funding from the Canada 150 Research Chairs Program. WJH is supported by a Royal Society University Research Fellowship. This research was enabled in part by support provided by Calcul Quebec (\url{www.calculquebec.ca}), WestGrid (\url{www.westgrid.ca}) and Compute Canada (\url{www.computecanada.ca})



\section*{Data Availability}
The data underlying this article will be shared on reasonable request to the corresponding author.

\bibliographystyle{mnras}
\bibliography{example} 




\appendix


\bsp	
\label{lastpage}
\end{document}